\documentclass[prd, aps,twocolumn,showpacs,floats,floatfix]{revtex4}
\usepackage[dvips]{graphicx}
\usepackage{amsmath}
\usepackage{subfigure}

\begin{document}
 
\title{Equilibrium Structure and Radial Oscillations of Dark Matter Admixed Neutron Stars}

\author{S.-C. Leung, M.-C. Chu, L.-M. Lin}
\affiliation{Department of Physics and 
Institute of Theoretical Physics, The 
Chinese University of Hong Kong, Hong Kong, China} 

\date{\today}

\begin{abstract} 
In [Leung {\it et al.}, Phys. Rev. D {\bf 84}, 107301 (2011)], 
we presented our results on using a general relativistic 
two-fluid formalism to study the hydrostatic equilibrium configuration of an 
admixture of degenerate dark matter and normal nuclear matter. 
In this work, we present more analysis to complement our previous findings. 
We study the radial oscillation modes of these compact stars in detail. 
We find that these stars in general have two classes of oscillation modes. 
For a given total mass of the star, the first class of modes is insensitive to 
the dark-matter particle mass. 
They also reduce properly to the oscillation modes of the corresponding 
ordinary neutron star, with the same total mass, when the mass fraction 
of dark matter tends to zero. On the other hand, the second class of modes
is due mainly to the dark-matter fluid. In the small dark-matter mass 
fraction limit, these modes are characterized purely by the oscillations 
of dark matter, while the normal matter is essentially at rest. 
In the intermediate regime where the mass fractions of the two fluids are 
comparable, the normal matter oscillates with the dark matter due to their
coupling through gravity. In contrast to the first class of modes, the frequencies of 
these dark-matter dominated modes depend sensitively on 
the mass of dark-matter particles. 
\end{abstract}


\maketitle{}

%%%%%%%%%%%%%%%%%%%%%%%%%%%%%%%%%%%%%%%%%%%%%%%%%%%%% 
\section{Introduction} 
\label{sec:intro}
%%%%%%%%%%%%%%%%%%%%%%%%%%%%%%%%%%%%%%%%%%%%%%%%%%%%% 

Dark matter (DM) has gained more support
for its existence from observations \cite{Roos:1},
such as galactic rotation curves \cite{Sofue:1, Catena:1, Weber:1}, 
stability of bars in spiral 
galaxies \cite{Miller:1, Hohl:1, Ostriker:1}, 
cosmological structure formation \cite{Springel2005}, and gravitational
lensing \cite{Massey:1}. 
However, the properties of DM particles, including their mass and 
interactions, are still largely unknown. 
Recent data from the DAMA \cite{DAMA2008} and 
CoGeNT \cite{CoGeNT2011} experiments are consistent with detecting 
light DM particles with mass $\sim 10$ GeV, which are incompatible with 
the null results from CDMS \cite{CDMS2011} and XENON \cite{XENON2010}.  
On the theoretical side, it has been suggested that 
isospin-violating DM may be the key to reconciling these experimental 
results \cite{Feng2011,Frandsen2011}.

Despite the uncertainties on DM properties, it is still interesting 
to ask how DM would affect stellar structure and whether one could in turn 
make use of stellar objects to put constraints on DM.
The role of DM in the first generation
stars as stellar seeds, together with the possibility of DM annihilation as 
the first phase in stellar evolution, are examined in 
\cite{Spolyar2008, Spolyar2009, Ripamonti2010, Hirano2011}.
The impacts of DM on the evolution and formation of main-sequence stars are
also examined \cite{Casanellas2009, Casanellas2011, Lopes2011}.  
The accretion and accumulation of non-self-annihilating DM particles
in the cores of planets in our solar system and some pulsars are proposed to 
cause changes in their orbits \cite{Peter2009, Iorio2010a, Iorio2010b}. 
Similarly, the effects of low-mass ($\sim 5$ GeV) asymmetric DM 
particles on the solar composition, oscillations, and neutrino fluxes have 
been considered recently \cite{Frandsen2010,Cumberbatch2010,Taoso2010}.

Besides main-sequence stars, one may infer DM particle properties through 
their effects on compact stars \cite{Goldman1989, Bertone:1, McCullough2010}. 
The effects of DM annihilation on the cooling curves of compact stars
are also studied in \cite{Kouvaris2008, Lavallaz2010, Kouvaris:2}. 
The response of neutron stars (NS) under non-self-annihilating DM models, 
such as asymmetric matter \cite{Kaplan:1} and mirror matter \cite{Saudin2009} 
have recently been studied. 
Neutron stars with DM cores are inherently two-fluid systems where the 
normal matter (NM) and DM couple essentially only through gravity. 
The technique used in recent studies of the structure of
these dark-matter admixed neutron stars (DANS) is based on an {\it ad hoc}
separation of the Tolman-Oppenheimer-Volkoff (TOV) equation into two different
sets for the normal and dark components inside the star
\cite{Saudin2009, Lavallaz2010,Ciarcelluti2011}.
This approach is motivated by the similarity of the structure equations
between the relativistic and Newtonian ones, but it is not derived from
first principle. 
In a previous paper \cite{Leung2011}, we made use of a general relativistic 
two-fluid formulation to study the equilibrium properties of 
DANS and proposed the existence of a new class of compact stars which are 
dominated by DM. These stars in general have a small NM core with radius
of a few kilometer embedded in a ten-kilometer-sized DM halo. 
In the present work, we describe the formulation of two-fluid DANS in 
greater detail. We present additional analysis to study the equilibrium 
properties and the radial oscillation modes of 
these stars in detail.

The outline of this paper is as follows: In Sec.~\ref{sec:formalism}
we present the relevant equations used to study the static equilibrium 
structure, moment of inertia and radial oscillation modes of DANS. 
In Sec.~\ref{sec:static_results} we study the static equilibrium properties of 
DANS. Sec.~\ref{sec:radial_results} discusses the stability of DANS 
and their radial oscillation modes in detail. 
Finally, we summarize our results in Sec.~\ref{sec:conclude}. 
We use units where $G=c=1$ unless otherwise noted.

%%%%%%%%%%%%%%%%%%%%%%%%%%%%%%%%%%%%%%%%%%%%%%%%%%%%% 
\section{Two-Fluid Formalism for DANS} 
\label{sec:formalism}
\subsection{Static equilibrium models}
%%%%%%%%%%%%%%%%%%%%%%%%%%%%%%%%%%%%%%%%%%%%%%%%%%%%%

The general relativistic two-fluid formalism was developed by 
Carter and his collaborators (see, e.g., \cite{Carter1989}). 
It has been employed in the study of superfluid neutron stars 
(e.g., \cite{Lee:1, Comer1999, Andersson2001, Prix:1}), where the two 
fluids are normal and superfluid nuclear matter. In this work, we adopt
the formulation in \cite{Comer1999} to study DANS. 
Here we shall briefly summarize the equations we used and refer 
the reader to \cite{Comer1999} for more details. 
To find the structure of a relativistic two-fluid star, one needs 
Einstein's equations $G_{\mu \nu}= 8 \pi T_{\mu \nu}$
coupled with a matter source. 
In the two-fluid formulation, the matter source is governed by the master
function $\Lambda (n^2, p^2, x^2)$, which is formed by the scalars
$n^2 = -n_\alpha n^\alpha$, $p^2 = - p_\alpha p^\alpha$, and 
$x^2 = - n_\alpha p^\alpha$. The four vectors $n^\alpha$ and $p^\alpha$
are the conserved NM and DM number density currents respectively\footnote{
In the original work for superfluid neutron stars \cite{Comer1999}, 
$n^\alpha$ stands for the superfluid-neutron number density currents, while 
$p^\alpha$ is the number density current for a conglomerate of all other 
charged constituents. }.
The master function is a two-fluid analog of the equation of state (EOS) and 
$-\Lambda$ is taken to be the thermodynamics energy density. 

For a static and spherically symmetric spacetime 
$ds^2=-e^{\nu(r)} dt^2 + e^{\lambda(r)} dr^2
+ r^2 (d\theta^2 + {\rm sin}^2 \theta d \phi^2)$, the Einstein equations 
and the fluid equations of motion reduce to the following differential
equations
\begin{eqnarray}
&&\lambda' = \frac{1 - e^{\lambda}}{r} - 8 \pi r e^{\lambda} \Lambda (n,p) ,   
\label{eq:metric_eq1}
\\
&&\nu' = - \frac{1 - e^{\lambda}}{r} + 8 \pi r e^{\lambda} \Psi (n,p) ,
\label{eq:metric_eq2}
\\
&&A^0_0 p' + B^0_0 n' + \frac{1}{2} (B n + A p) \nu ' = 0,  
\label{eq:fluid_eq1}
\\
&&C^0_0 p' + A^0_0 n' + \frac{1}{2} (A n + C p) \nu ' = 0,
\label{eq:fluid_eq2}
\end{eqnarray} 
where the primes refer to derivative with respect to $r$ and
\begin{equation}
A = - {\partial \Lambda \over  \partial x^2} , \ \ 
B = -2 {\partial \Lambda \over \partial n^2} , \ \ 
C = - 2 {\partial \Lambda \over \partial p^2} .
\end{equation}
The coefficients $A^0_0$, $B^0_0$, and $C^0_0$ are given by 
\begin{eqnarray}
&&A^0_0 = A + 2 \frac{\partial B}{\partial p^2} np + 2 \frac{\partial 
A}{\partial n^2} n^2 + 2
\frac{\partial A}{\partial p^2} p^2 + \frac{\partial A}{\partial x^2} pn,
\\
&&B^0_0 = B + 2 \frac{\partial B}{\partial n^2} n^2 + 4 \frac{\partial 
A}{\partial n^2} n p +
\frac{\partial A}{\partial x^2} p^2,
\\
&&C^0_0 = C + 2 \frac{\partial C}{\partial p^2} p^2 + 4 \frac{\partial 
A}{\partial p^2} n p +
\frac{\partial A}{\partial x^2} n^2.
\end{eqnarray}
The generalized pressure $\Psi$ in Eq.~(\ref{eq:metric_eq2}) is computed by 
\begin{equation}
\Psi (n, p) = \Lambda + \mu n + \chi p , 
\end{equation}
where $\mu = B n + A p$ and $\chi = C p + A n$. 

Using Eqs. (\ref{eq:metric_eq1})-(\ref{eq:fluid_eq2}), one may calculate 
numerically the structure of a star by choosing suitable boundary
conditions, namely the central densities of the two fluids
$n_c$ and $p_c$, $\lambda(0)= \lambda '(0)=0$ and $\nu '(0)=0$.
$\nu(0)$ is fixed by matching the solution with
the Schwarzschild metric at the star surface $r=R$.
The surface of the NM at $r=R_{\rm NM}$ is defined by the condition
$n(R_{\rm NM})=0$. Similarly, we have the condition $p(R_{\rm DM})=0$ to 
define the surface of the DM fluid at $r=R_{\rm DM}$. 
In practice we choose $n$ or $p = 10^{-5}$ fm$^{-3}$ to define the surface of 
either fluid. 
It should be noted that in general the two surfaces are different 
(i.e., $R_{\rm NM} \neq R_{\rm DM}$). The star surface $r=R$, where a 
matching to the Schwarzschild metric is performed, is defined to be the larger
one of $R_{\rm NM}$ and $R_{\rm DM}$. 
The total mass of the star is computed by
\begin{equation}
M = -4 \pi \int^R_0 dr r^2 \Lambda(r),
\end{equation}
while the total particle masses (baryonic masses) of NM and DM are computed by
\begin{eqnarray}
M_{NM} = 4 \pi m_{n} \int^{R_{\rm NM}}_0 dr r^2 e^{\lambda/2} n, \\
M_{DM} = 4 \pi m_{{\rm DM}}\int^{R_{\rm DM}}_0 dr r^2 e^{\lambda/2} p,
\end{eqnarray}
where $m_{n}$ and $m_{DM}$ are the particle masses
of NM and DM respectively. It should be noted that the sum of the baryonic 
masses $M_{\rm NM}+M_{\rm DM}$ is in general different from the total mass $M$,
which also includes the contributions from gravitational and internal 
energies. 

%%%%%%%%%%%%%%%%%%%%%%%%%%
\subsection{Choice of EOS}
%%%%%%%%%%%%%%%%%%%%%%%%%%%

In the two-fluid formalism, the master function $\Lambda$ plays the role of 
the EOS information needed in the structure calculation. 
In this work, we assume that DM couples with NM only through
gravity. Hence, the master function does not depend on the scalar product
$x^2= - n_\alpha p^\alpha$ and is separable in the sense that 
\begin{equation}
\Lambda(n,p) = \Lambda_{\rm NM}(n) + \Lambda_{\rm DM} (p) ,  
\end{equation}
$\Lambda_{\rm NM} (n)$ and $\Lambda_{\rm DM} (p)$ being the negative of energy
densities of NM and DM at a given number density,  respectively.
These assumptions imply that the coefficients $A=A^0_0=0$ in our study. 

We choose the APR EOS \cite{Akmal:1} for NM (and BBB2 EOS \cite{Baldo1997} as 
well for comparison) and ideal degenerate Fermi gas EOS for DM. 
As discussed earlier, DM candidates in the mass range of a few GeV are of 
great interest recently.
We shall thus consider fermionic DM particles in this mass range.

%%%%%%%%%%%%%%%%%%%%%%%%%%%%%%%%%%%%%%%%%%%%%%%%%%%%%%%%%%%
\subsection{Moment of inertia}
%%%%%%%%%%%%%%%%%%%%%%%%%%%%%%%%%%%%%%%%%%%%%%%%%%%%%%%%%%%

Besides global quantities like gravitational mass and radius, it is also 
interesting to study the moment of inertia of DANS since it is measurable 
and plays an important role in the physics of ordinary neutron stars. 

The moment of inertia of a rotating star in general relativity is defined 
by $I = J/\Omega$ in the slow rotation limit, where $J$ and $\Omega$ are 
respectively the angular momentum and angular velocity. 
Here we follow the formulation developed in \cite{Andersson2001} to calculate
the angular momentum of a two-fluid star in the slow rotation limit. 
First, we need to compute the frame-dragging of the star due to its rotation. 
In general, DM and NM can rotate with different velocities,
$\Omega_p$ and $\Omega_n$ respectively. 
The frame-dragging $\omega$ is given by \cite{Andersson2001}
\begin{equation}
\begin{split}
\frac{1}{r^4} (r^4 e^{-(\lambda+\nu)/2} L_n')' - 16 \pi e^{(\lambda - \nu)/2} 
(\Psi_0 - \Lambda_0) L_n = \\
16 \pi e^{(\lambda - \nu)/2} \chi_0 p_0 (\Omega_n - \Omega_p),
\label{eq:drag_eq}
\end{split}
\end{equation}
where
\begin{equation} 
L_n = \omega - \Omega_n,
\end{equation}
\begin{equation}
L_p = \omega - \Omega_p.
\end{equation}
It should be noted that Eq. (\ref{eq:drag_eq}) is formally identical 
to the equation obtained by Hartle \cite{Hartle:1} for one-fluid stars
except for the nonzero source term on the right-hand side. 
In particular, the corotating case ($\Omega_n = \Omega_p$) reduces to 
the one-fluid result.
To integrate this equation with higher accuracy, one defines a new variable
\begin{equation}
T_n = (r^4 e^{-(\lambda+\nu)/2} L_n').
\end{equation}
We can decompose Eq. (\ref{eq:drag_eq}) into two
first-order equations: 
\begin{equation}
\begin{split}
T_n' =  16 \pi r^4 e^{(\lambda - \nu)/2}
(\Psi_0 - \Lambda_0) L_n + \\
16 \pi r^4 e^{(\lambda - \nu)/2} \chi_0 p_0 (\Omega_n - \Omega_p),
\end{split}
\end{equation}
\begin{equation}
L_n' = \frac{1}{r^4} e^{(\lambda+\nu)/2} S_n.
\end{equation}
We integrate
the two variables $T_n$ and $L_n$ from the origin
to the surface subject to the boundary condition
\begin{equation}
L_n(R) = -\Omega_n - \frac{R}{3} \left( \frac{d L_n}{dr} \right) _{r=R}.
\end{equation}

With $L_n$ and $L_p$ calculated above, the total angular momentum of the star
is given by 
\begin{equation}
J = - \frac{8 \pi}{3} \int^R_0 dr r^4 e^{(\lambda - \nu)/2} (\mu_0 n_0 L_n 
+ \chi_0 p_0 L_p).
\label{eq:moi}
\end{equation}
When the master function $\Lambda$ is independent of the scalar product 
$x^2=-n_\alpha p^\alpha$ (as we assume in this work), the two terms in the integral 
correspond to the NM and 
DM angular momenta ($J_{n}$ and $J_{p}$) respectively 
\cite{Andersson2001}. 
We can thus define the moments of inertia 
$I_{n}=J_{n}/\Omega_{n}$ and
$I_{p}=J_{p}/\Omega_{p}$.
$I_{n}$ and $I_{p}$ depend on the nonrotating background
quantities of NM and DM separately. The total moment of inertia of DANS
is then defined by $I = I_{n} + I_{p}$.

%%%%%%%%%%%%%%%%%%%%%%%%%%%%%%%%%%%%%%%%%%%%%%%%%%%%%%%%%%%%%%%%%%%%
\subsection{Equations for radial oscillations}
%%%%%%%%%%%%%%%%%%%%%%%%%%%%%%%%%%%%%%%%%%%%%%%%%%%%%%%%%%%%%%%%%%%

To study the stability and radial oscillation modes of DANS, we use the set of 
equations for radial perturbations of a two-fluid star derived in Sec. IV-B of
\cite{Comer1999}. We shall discuss our numerical scheme in solving the 
eigenvalue problem in detail.

Assuming DM has no interaction with NM except through gravity (in the sense 
that the coefficient $A = A^0_0 = 0$ ), 
and adopting the notation $S_n = r n W_n$ and $S_p = r p W_p$, the 
set of equations governing radial oscillations is simplified to

\begin{equation}
\begin{split}
\frac{\omega^2}{r^2} e^{(\lambda - \nu)/2} (B S_n) =
\large( -\frac{e^{(\nu - \lambda)/2}}{r^2} [B^0_0 S_n']+ \\
\frac{1}{2} \frac{e^{\nu/2}}{r} [B^0_0 n]  [8 \pi e^{\lambda /2} (\chi
S_p + \mu S_n)] \large)' + \frac{1}{2} \mu H_0',   
\label{eq:radial_oscillation_1}
\end{split}
\end{equation}
\begin{equation}
\begin{split}
\frac{\omega^2}{r^2} e^{(\lambda - \nu)/2} (C S_p) =
\large( - \frac{e^{(\nu - \lambda)/2}}{r^2} [C^0_0 S_p']+ \\
\frac{1}{2} \frac{e^{\nu/2}}{r} [C^0_0 p]  [8 \pi e^{\lambda /2} (\chi
S_p + \mu S_n)] \large)' + \frac{1}{2} \chi H_0',
\label{eq:radial_oscillation_2}
\end{split}
\end{equation}
where
\begin{equation}
\begin{split}
H_0' = 4 \pi r e^{\nu/2 + \lambda} \large( p^2 C^0_0 + n^2 B^0_0 - 2 \Psi - \frac{1}{4 \pi r^2} \large) H_2 \\
-  \frac{8 \pi e^{(\nu + \lambda) /2}}{r} \large( p C^0_0 S_p' + n B^0_0 S_n' \large),
\end{split}
\end{equation}
\begin{equation}
H_2 = \frac{8 \pi e^{\lambda /2}}{r} (\chi S_p + \mu S_n).
\end{equation}  
Note that the quantities $W_n$ and $W_p$
are related to the radial component of the 
Lagrangian displacement of both fluids 
by $\delta \xi^r_n =  e^{-\lambda/2} W_n e^{i \omega t} /r$
and $\delta \xi^r_p =  e^{-\lambda/2} W_p e^{i \omega t} /r$.

The radial oscillation modes can be solved by specifying
the boundary conditions at the core, and the correct eigenvalues
can be obtained by checking whether the boundary conditions at
the surfaces of both fluids are satisfied. In this problem,
we have 4 degrees of freedom at the core, namely
$S_n(0)$, $S_p(0)$, $S_n'(0)$ and $S_p'(0)$. 
Note that the four variables cannot be 
set arbitrarily, otherwise the boundary
conditions of both fluids at the surfaces
cannot be satisfied simultaneously. 

The boundary condition at the surface of each fluid is given by the vanishing 
of the Lagrangian density variation of the fluid.  
The Lagrangian density variations are given by \cite{Comer1999}
\begin{equation}
\frac{\Delta n}{n} = e^{-\lambda/2} \left( \frac{W_n}{r^2} +
\frac{W_n'}{r} \right) - \frac{1}{2} H_2,  
\label{eq:delta_n_over_n}
\end{equation}
\begin{equation}
\frac{\Delta p}{p} = e^{-\lambda/2} \left( \frac{W_p}{r^2} +
\frac{W_p'}{r} \right) - \frac{1}{2} H_2 . 
\label{eq:delta_p_over_p}
\end{equation}
Hence, we require the boundary conditions $\Delta n/n = 0$ at $r = R_{NM}$
and $\Delta p/p = 0$ at $r = R_{DM}$.

To find the correct boundary conditions at the core, 
we first choose $S_n$ to be unity.
$S_n'(0)$ is chosen according to the 
Taylor expansion of $S_n(r)$ 
and the regularity condition of $S_n$ near the center.
Namely, we have 
\begin{equation}
S_n(r) = S_n(0) r^3 + O(r^5),
\end{equation}
from which we have $S_n'(0) = 3 S_n(0)$ [similarly, we have  
$S_p'(0) = 3 S_p(0)$]. $S_p(0)$ is chosen such that the boundary condition 
of the inner surface of a DANS model can be satisfied. 
In practice, we first choose a trial eigenvalue $\omega$
and a trial $S_p(0)$ (without loss of 
generality, we assume $R_{DM}<R_{NM}$ in this discussion). 
Then, we integrate up to the inner surface of the DANS model. 
If the boundary condition at the inner surface is not satisfied, 
a new trial of $S_p(0)$ is used and the integration is repeated. 
Once a correct trial of $S_p(0)$ is found to satisfy the boundary 
condition at the inner surface, we continue the integration up to the 
outer surface and check whether the boundary condition at the 
outer surface is satisfied. We obtain the eigenvalue if the trial 
$\omega$ can satisfy the boundary condition of the outer surface. 
%In short, we have only one degree of freedom $S_n(0)$,
%and the amplitude of $S_p(0)$ is fixed once $S_n$ is chosen,
%while $S_n'(0)$ and $S_p'(0)$ are determined by the first
%non-zero terms in the Taylor expansion of both $S_n(r)$ and 
%$S_p(r)$.

\begin{figure}
\centering
\subfigure[]{
\includegraphics[width=8cm, height=6cm]{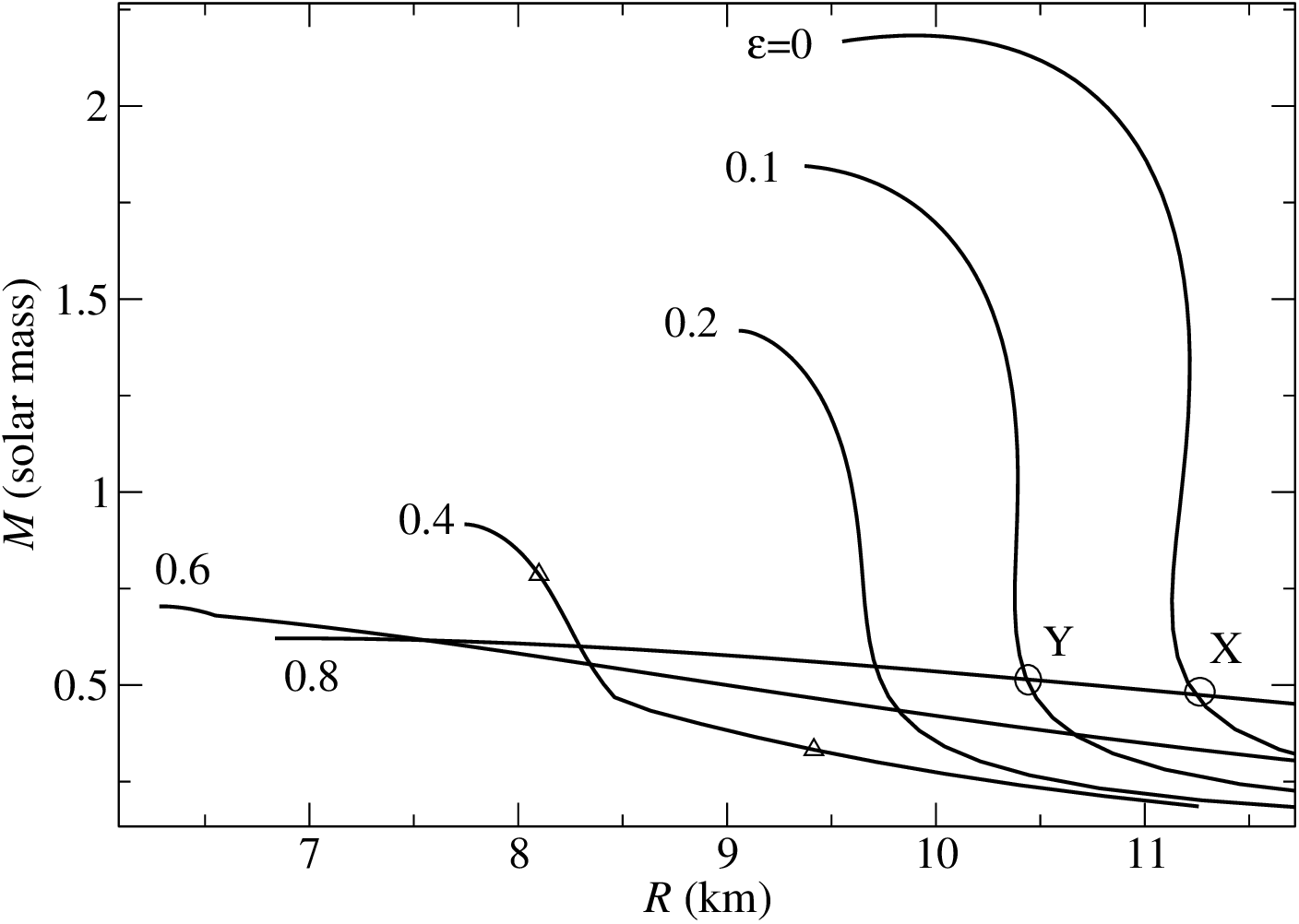}
\label{fig:MR_relation}
}
\\
abc
\\
\subfigure{
\includegraphics[width=8cm, height=6cm]{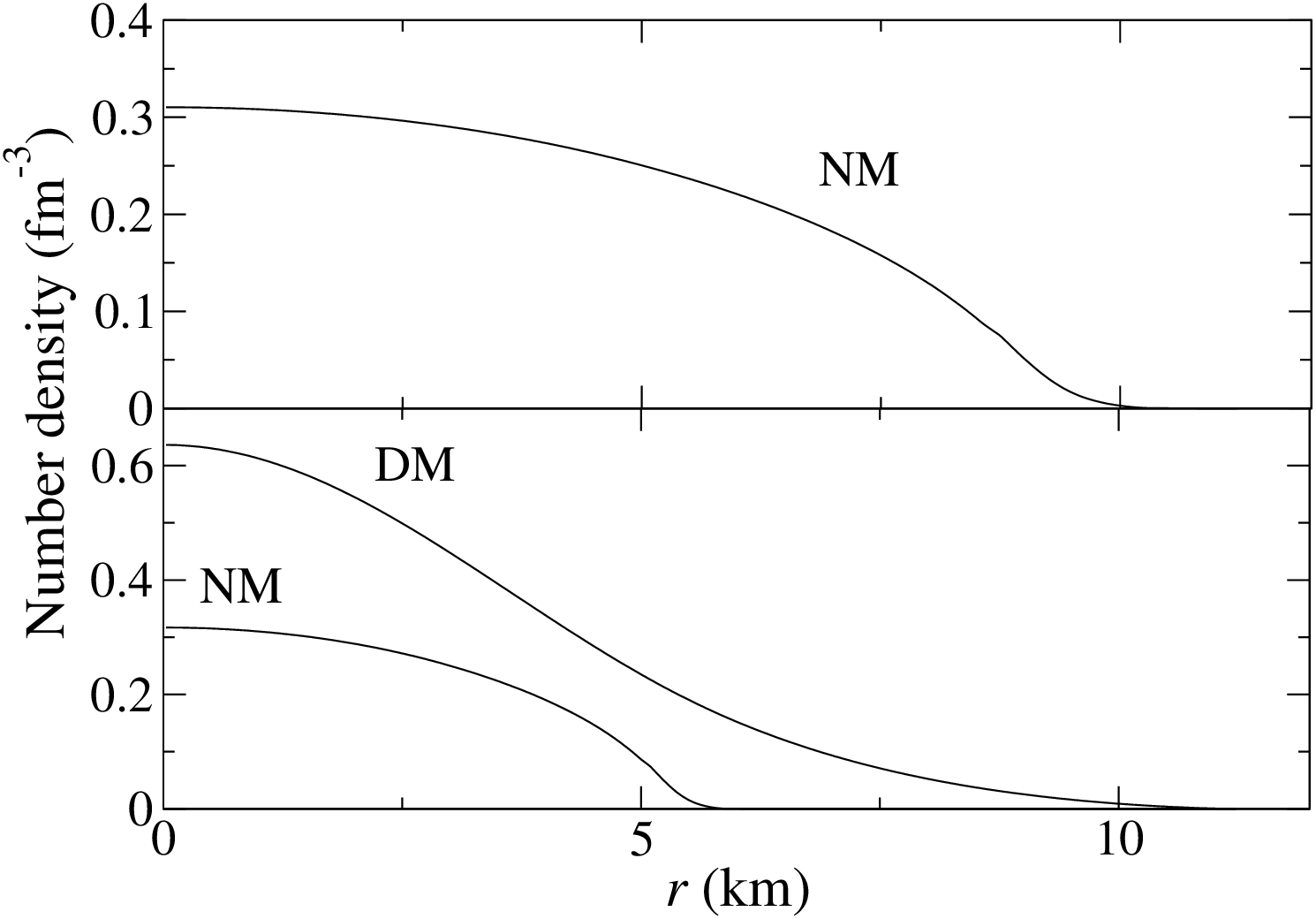}
\label{fig:profile2}
}
\caption{Upper plot (a): Mass-radius relations of DANS for different
DM mass fraction $\epsilon$. The NM is modeled by the APR EOS and the 
DM particle mass is 1 GeV. Masses are in unit of $M_{\odot}$.
Lower plot (b):
Density profiles of the star models at the point X in
Fig. \ref{fig:MR_relation} for an ordinary NS
($\epsilon = 0$, upper panel) and a DM
dominated star ($\epsilon = 0.8$, lower
panel). Both star models have mass 0.475
$M_{\odot}$ and $R = 11.25$ km. For the DM dominated star (lower panel), 
$M_{NM} = 0.0981 M_{\odot}$, $M_{DM}=0.3923 M_{\odot}$ and $R_{NM} = 6.29$ km.}
\end{figure}

\begin{figure}
\centering
\includegraphics[width=8cm, height=6cm]{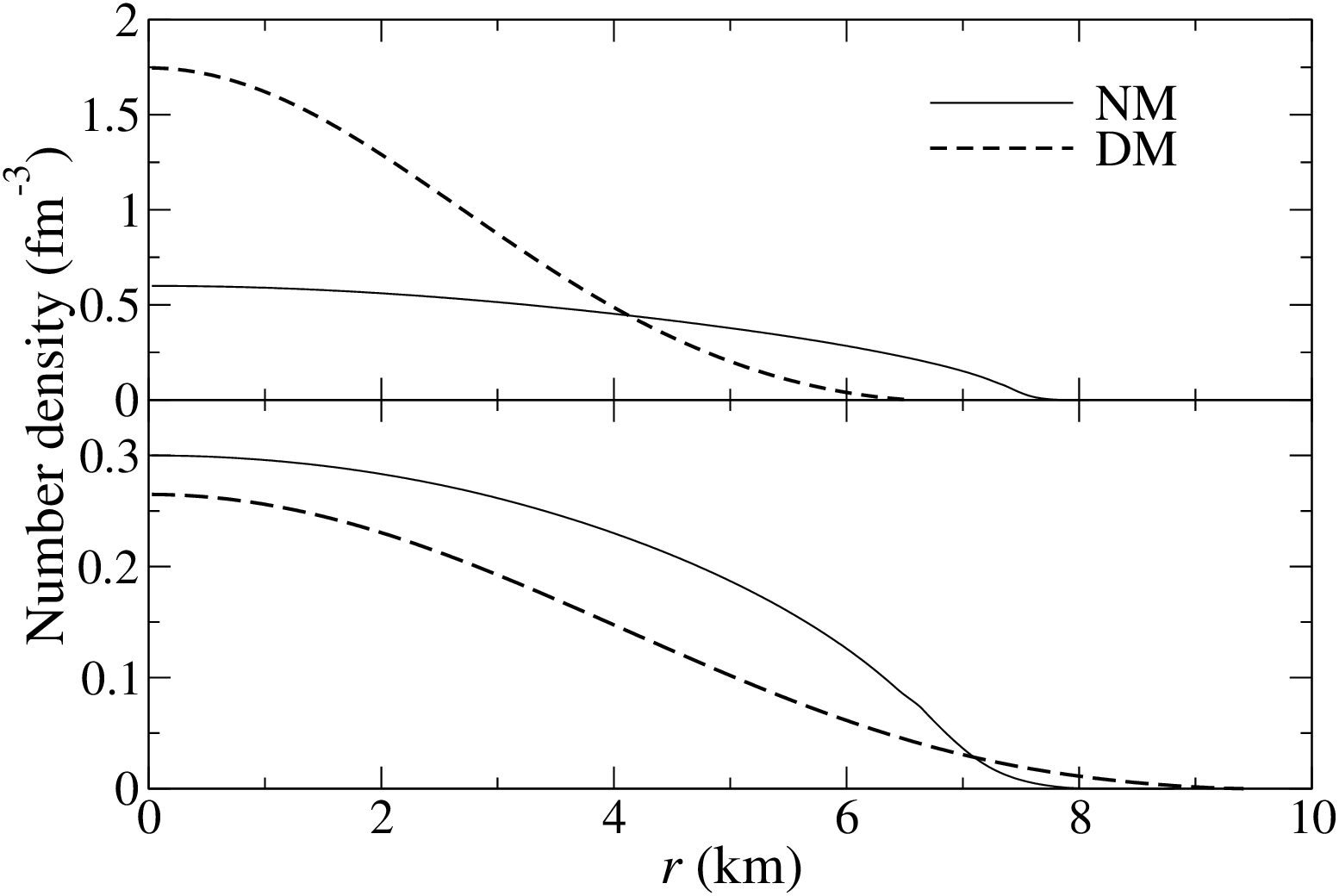}
\caption{Density profiles for two star models with $\epsilon$ = 0.4
with the APR EOS for NM and ideal degenerate gas EOS for DM
with DM particle mass = 1 GeV. 
The solid (dashed) lines are for NM (DM).  
Upper plot: $M=0.780 M_{\odot}$, $R=8.10$ km, $R_{DM} = 6.60$ km.
Lower plot: $M=0.332 M_{\odot}$, $R=9.42$ km, $R_{NM} = 8.84$ km. }
\label{fig:density_profile_04}
\end{figure}

%%%%%%%%%%%%%%%%%%%%%%%%%%%%%%%%%%%%%%%%%%%%%%%%%%%%%
\section{Static Equilibrium properties of DANS}
\label{sec:static_results}
%%%%%%%%%%%%%%%%%%%%%%%%%%%%%%%%%%%%%%%%%%%%%%%%%%%%%

\subsection{General Properties of DANS}

In Fig. \ref{fig:MR_relation}, we show the mass-radius 
relations of DANS for different DM mass fractions defined by 
\begin{equation}
\epsilon  = { M_{DM} \over M_{NM} + M_{DM} } . 
\end{equation}
The NM is modeled by the APR EOS and the DM particle mass is 
$m_{\rm DM}=1$ GeV.

The case $\epsilon=0$ corresponds to ordinary NS without DM. 
For increasing $\epsilon$, representing DANS with higher DM proportion, 
we observe two results: first, the maximum stable mass decreases and the 
stars also have smaller radii. 
For example, for $\epsilon = 0.2$, the mass and radius of the maximum 
stable mass configuration are decreased by 35\% and 9\% respectively,
compared to the case with $\epsilon = 0$.
Second, the $M$-$R$ curve flattens as $\epsilon$ increases, such as the one 
shown for $\epsilon$ = 0.8.
For intermediate DM mass fraction, such as $\epsilon=0.4$, the curves are 
made of two parts: The curve is flat at large $R$, but at small $R$, it is 
qualitatively similar to the curve for ordinary NS. 
The above patterns suggest that a new class of compact stars exists, when 
$\epsilon$ is sufficiently large.
We will show that these DM dominated stars are 
qualitatively different from ordinary NS in many aspects. 

We plot in Fig. \ref{fig:profile2}
the density profiles of the DANS models marked
as point X in Fig. \ref{fig:MR_relation}. The upper 
panel is for the case $\epsilon=0$ while the lower
one is for the case $\epsilon = 0.8$. 
Although both models have the same total mass and radius, 
the mass in the upper one is contributed only by NM, 
while the mass in the lower one is mainly contributed by DM. 
For the case $\epsilon=0.8$, it is seen that a small NM core is embedded in a 
larger DM halo. 

Next we plot in Fig. \ref{fig:density_profile_04} two distinctive models
for the case $\epsilon = 0.4$ [marked with triangles
in Fig. \ref{fig:MR_relation}]. One star is chosen from 
the part of the $M$-$R$ curve which is similar to that of ordinary NS, 
while another star is chosen from the other side. We choose one model 
(upper panel in Fig. \ref{fig:density_profile_04}) to have
$M=0.780 M_{\odot}$, $R= 8.10$ km, $R_{DM} = 6.60$ km and the other 
(lower panel) to have $M=0.332 M_{\odot}$, $R =9.42$ km, $R_{NM} = 8.84$ km.
For the former model (upper panel), we see that the DM is engulfed by NM. 
However, the situation is reversed for the latter model (lower panel). 

\begin{figure}
\centering
\includegraphics[width=8cm, height=6cm]{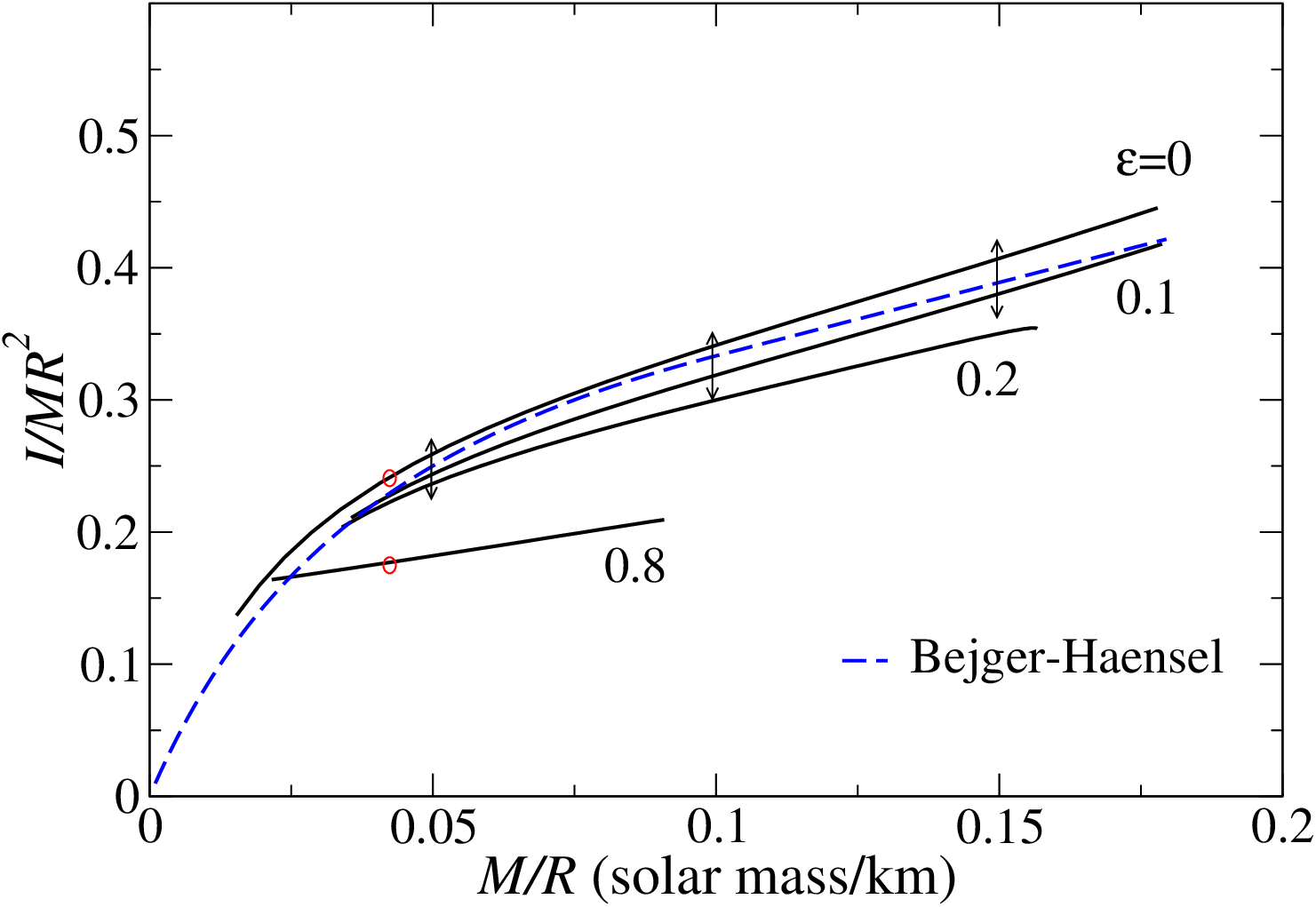}
\caption{$I/MR^2$ of DANS against $M/R$ with the APR EOS
for NM and ideal degenerate Fermi gas EOS for DM,  
with DM particle mass of 1 GeV. The dashed line
corresponds to the numerical fitting found by
Bejger and Haensel \cite{Haensel:1}.
The circles correspond to the two stellar models
($\epsilon = 0$ and 0.8) at the point X in Fig. \ref{fig:MR_relation}}
\label{fig:inertia_compact_APR}   
\end{figure}

The scaled moment of inertia $\tilde{I} \equiv I/MR^2$ of DANS
is plotted as a function of the compactness ($M/R$, in the unit solar mass/km) 
in Fig. \ref{fig:inertia_compact_APR}.
In \cite{Haensel:1}, Bejger and Haensel found an approximate universal 
relation between $\tilde{I}$ and compactness
($z \equiv (M/M_{\odot})/(\rm{km}/{\it R})$):
\begin{eqnarray}
\tilde{I} = \frac{z}{0.1 + 2z}, ~~ z<0.1,  
\label{eq:I_universal_1}
\\
\tilde{I} = \frac{2}{9}(1 + 5z), ~~ z>0.1. 
\label{eq:I_universal_2}
\end{eqnarray}
This formula is shown as the dashed line in Fig.~\ref{fig:inertia_compact_APR}.
The vertical lines (with arrows) at $z=0.05$, 0.1 and 0.15 represent the 
range of values of $\tilde{I}$ obtained by a large set of EOS models which 
were used to obtain the formula.
They can be regarded as the error bars of 
Eqs. (\ref{eq:I_universal_1}) and (\ref{eq:I_universal_2}) at those values of 
$M$/$R$. 
The circles in the figure correspond to an ordinary neutron star
($\epsilon = 0$) and DM dominated star ($\epsilon = 0.8$) at the point X
in Fig. \ref{fig:MR_relation}. 
While the scaled moment of inertia of ordinary
neutron stars can be modeled approximately by Eq. (\ref{eq:I_universal_1})
and (\ref{eq:I_universal_2}),
Fig. \ref{fig:inertia_compact_APR} shows that 
${\tilde I}$ of DANS depends sensitively on the
amount of DM. In particular, for the DM dominated sequence
$\epsilon = 0.8$, the value of ${\tilde I}$ is significantly smaller
than that allowed for ordinary neutron stars with the
same compactness.

%%%%%%%%%%%%%%%%%%%%%%%%%%%%%%%%%%%%%%%%%%%%%%%%%%%%%
\subsection{Linear response of DANS}
%%%%%%%%%%%%%%%%%%%%%%%%%%%%%%%%%%%%%%%%%%%%%%%%%%%%%

Here we study the effects of DM in the core on the structure of DANS at 
the small $M_{\rm DM}$ limit.  
In the linear regime, namely a DANS with small $M_{\rm DM}$, the radius and 
moment of inertia of the star vary linearly with $M_{\rm DM}$. 
But the slope of the linear variation depends on the stiffness of the NM EOS.
The stiffer the NM EOS is, the smaller the contraction results. 

\begin{figure}
\centering
\subfigure[]{
\includegraphics[width=8cm, height=6cm]{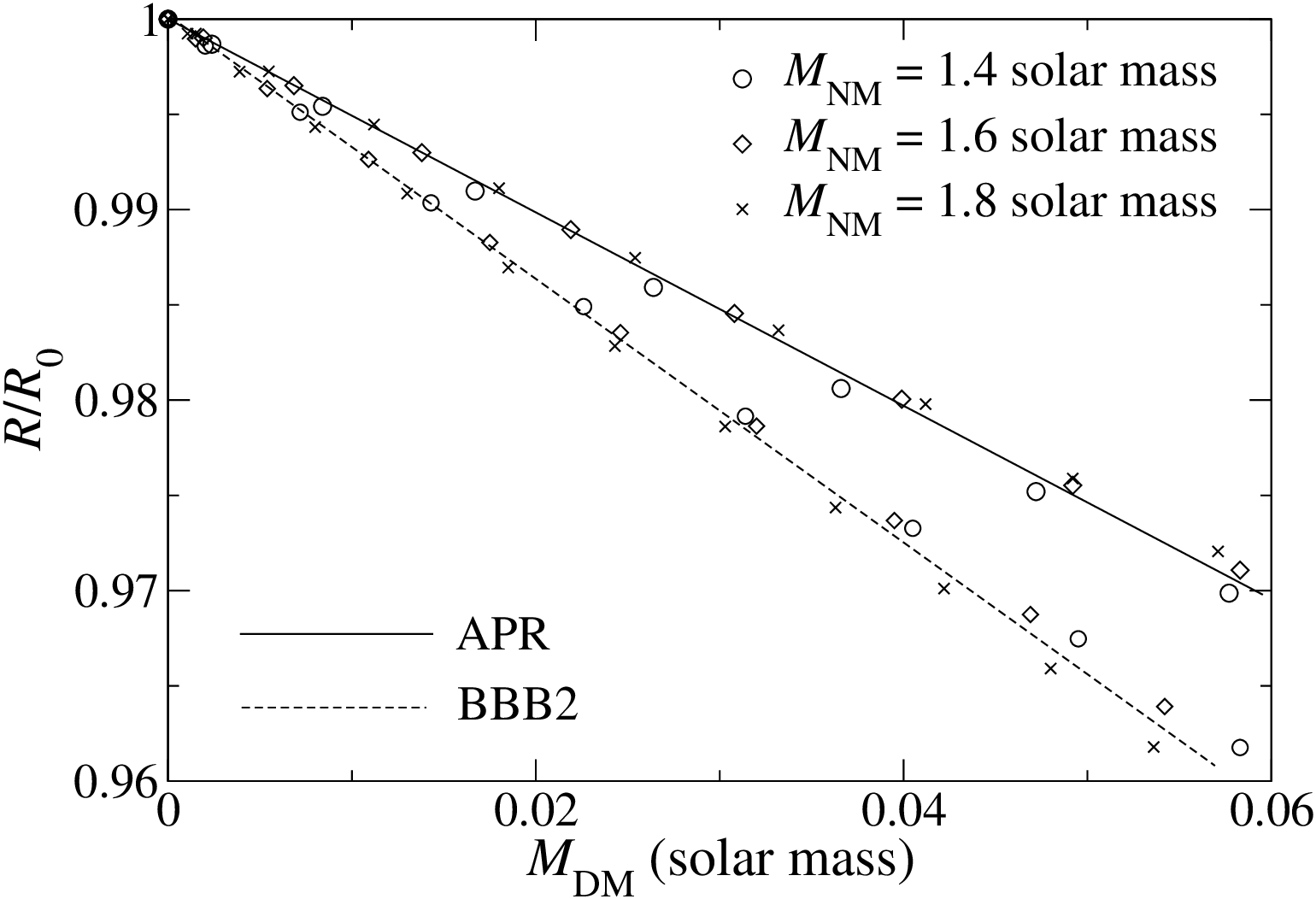}
\label{fig:dmmass_radius}
}
\\
abc
\\
\subfigure{
\includegraphics[width=8cm, height=6cm]{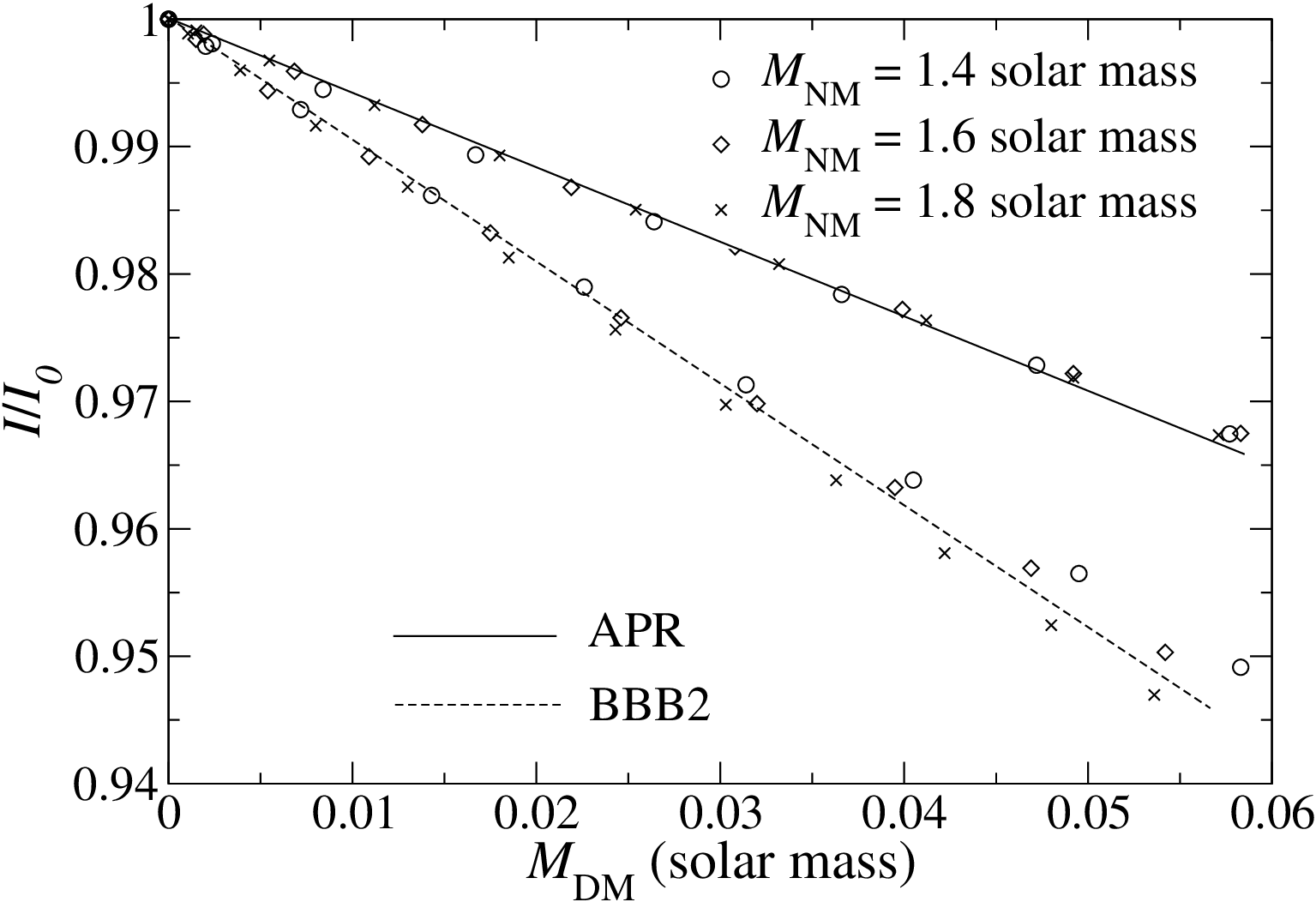}
\label{fig:dmmass_inertia}
}
\caption{Upper plot (a): $R/R_0$ is plotted against $M_{\rm DM}$
(in $M_{\odot}$) for the APR and BBB2 EOS for NM and ideal
degenerate Fermi gas EOS for DM, with DM particle mass of 1 GeV.
$R_0$ is the NS radius without DM.
Lower plot (b): $I/I_0$ against $M_{\rm DM}$ (in $M_{\odot}$).
$I_0$ is the moment of inertia without DM.}
\end{figure}

In Fig. \ref{fig:dmmass_radius} we plot 
$R/R_0$ against $M_{\rm DM}$, where $R_0$ is the radius  
without DM (i.e., an ordinary NS), for $M_{\rm NM}$ = 1.4, 1.6, and 
1.8 $M_{\odot}$. We see that the relative changes in radii are
linear in $M_{\rm DM}$. For a fitting in the form
\begin{equation}
\frac{R}{R_0} = 1 + a_1 M_{DM},
\end{equation}
we find that $a_1 \approx -0.5$ for the APR EOS and 
$a_1 \approx -0.7$ for the BBB2 EOS. The larger magnitude of $a_1$ for BBB2 
EOS is due to the fact that this EOS is softer than the APR EOS. 
We also see that $a_1$ is essentially independent of $M_{\rm NM}$.
The effect from DM is no longer linear for large $M_{\rm DM}$.

\begin{figure} 
\centering 
\subfigure[]{
\includegraphics[width=8cm, height=6cm]{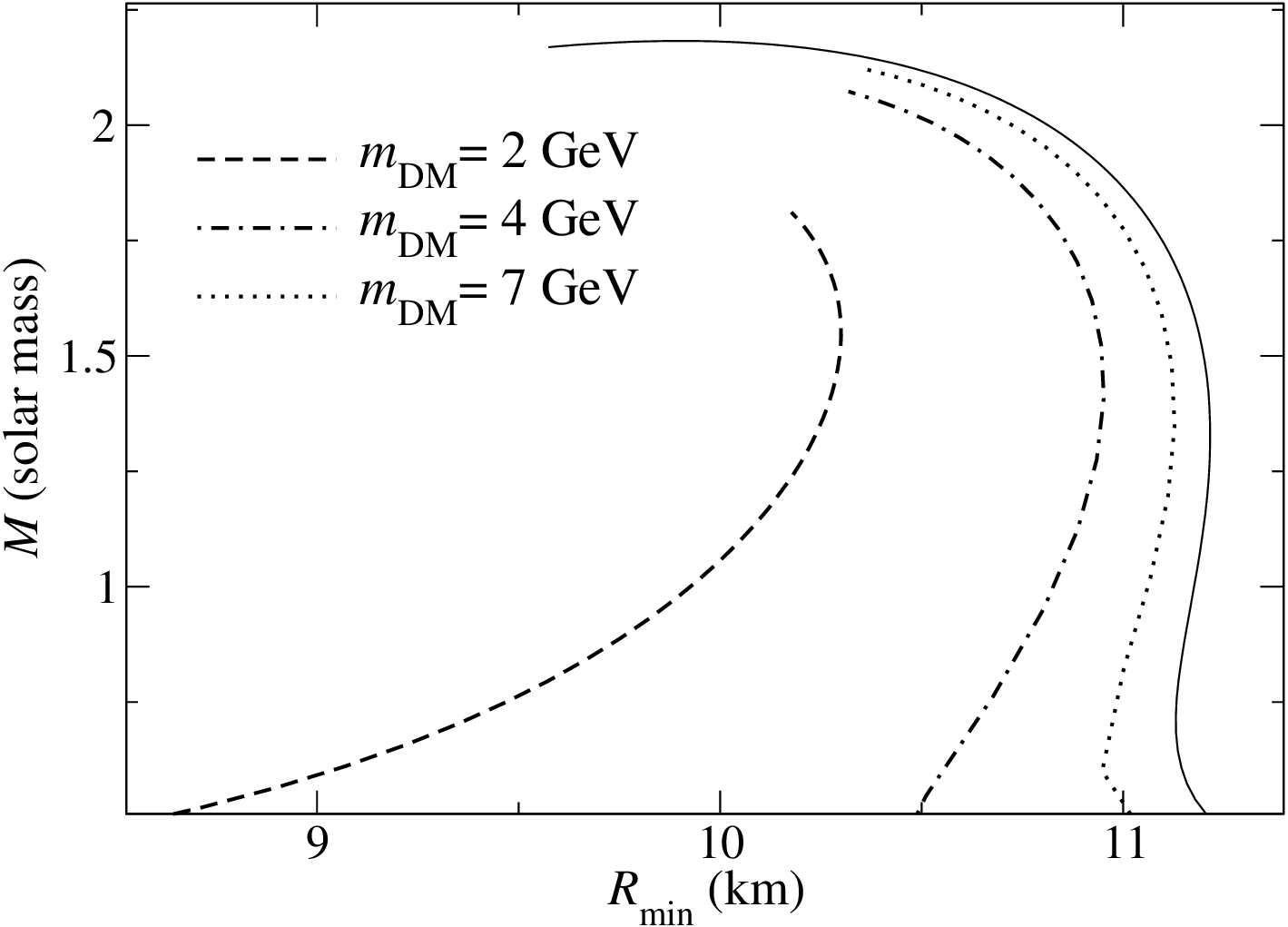}
\label{fig:MR_mass}
}
\\
abc
\\
\subfigure{
\includegraphics[width=8cm, height=6cm]{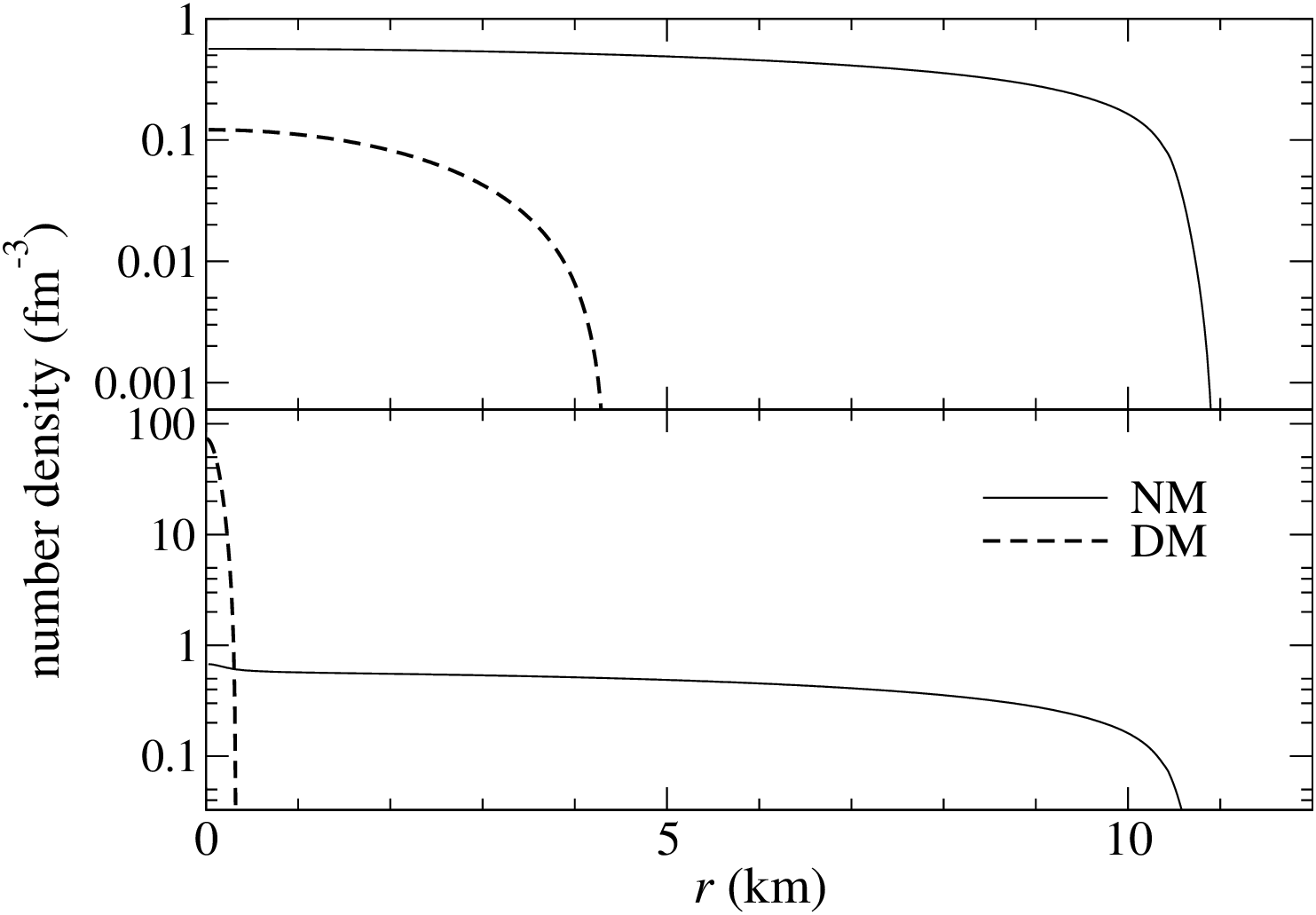}
\label{fig:density_profile_mass}
}
\caption{Upper plot (a): Total mass $M$ is plotted against the minimum
radius $R_{\rm min}$ of stable stars allowed for a given DM particle 
mass $m_{\rm DM}$. The solid line is the $M$-$R$ curve for ordinary NS 
without DM for comparison. 
Lower plot (b): Number density profiles of NM (solid lines) and DM 
(dashed lines) for two stars with the same $M = 1.4 M_{\odot}$ and 
$M_{DM} = 0.01 M_{\odot}$, but with $m_{\rm DM}=1$ GeV (upper panel) and 
7 GeV (lower panel).  }
\end{figure}

In Fig. \ref{fig:dmmass_inertia} we plot $I/I_0$ against $M_{\rm DM}$, 
where $I_0$ is the moment of inertia of a NS without DM, for three different 
values of $M_{\rm NM}$ as above. 
The change in $I$ is almost the same for the three cases:
\begin{equation}
\frac{I}{I_0} = 1 + a_2 M_{DM},
\label{eq:linear_response_inertia}
\end{equation}
with $a_2 \approx -0.6$ for the APR EOS and $a_2 \approx -0.9$ for the BBB2 
EOS. Again, the value of $a_2$ does not depend strongly on $M_{\rm NM}$.

%\begin{figure}
%\centering
%\includegraphics[width=8cm, height=6cm]{fig6.eps}
%\caption{Relative change of radius of a DANS with 
%the APR EOS for the NM and the ideal degenerate gas EOS for DM for a given
%$M_{DM}$ but with different $m_{DM}$. $R_0$ stands for
%the radius when no DM is present.}
%\label{fig:radius_ratio_mass}
%\end{figure}

%%%%%%%%%%%%%%%%%%%%%%%%%%%%%%%%%%%%%%%%%%%%%%%%%%%%%%%%%%%%%%%%%%%
\subsection{Effects of DM particle mass}
%%%%%%%%%%%%%%%%%%%%%%%%%%%%%%%%%%%%%%%%%%%%%%%%%%%%%%%%%%%%%%%%%%

As discussed in Sec.~\ref{sec:intro}, DM candidates in the mass range a few 
GeV is of great interest recently. We shall thus study the 
effects of DM particle mass $m_{\rm DM}$ in this range. 
We shall compare the equilibrium properties of DANS with different 
$m_{\rm DM}$. In this part, we use the APR EOS to model the NM. 

First, we plot in Fig.~\ref{fig:MR_mass} the total mass $M$ against the 
minimum radius $R_{\rm min}$ of stable stars allowed for a 
given $m_{\rm DM}$. We present the results for three different cases 
$m_{\rm DM}=1$, 4, and 7 GeV. We also plot the $M$-$R$ curve (solid line) 
for ordinary NS without DM for comparison.  
For fixed $M$ and $m_{\rm DM}$, the radius of the star decreases as the 
mass fraction of DM increases (i.e., the star becomes more compact). 
We define $R_{\rm min}$ to be the minimum radius below which the star becomes 
unstable. For example, Fig.~\ref{fig:MR_mass} shows that the minimum radius 
allowed for a DANS with total mass $M=1 M_\odot$ and DM particle mass 
$m_{\rm DM}=2$ GeV is about 10 km. We also see that $R_{\rm min}$ increases 
with $m_{\rm DM}$ for a given $M$. 
%Notice that in Fig. \ref{fig:MR_mass} each curve ends at the point 
%where the density for NM reaches the maximum density of the EOS table.
%We choose not to extrapolate the EOS due to further complication of 
%uncertainties in nuclear matter properties at ultra-high density,
%such as the transition to quark matter or creation of exotic matter.

In Fig. \ref{fig:density_profile_mass}, we plot the density profiles
of two stars with the same $M = 1.4 M_{\odot}$ and $M_{DM} = 0.01 M_{\odot}$, 
but with different DM particle mass $m_{DM} = 1$ GeV (upper panel) and 
7 GeV (lower panel). In Fig. \ref{fig:density_profile_mass},
we see that with a higher DM particle mass $m_{\rm DM} = 7$ GeV, the DM core 
shrinks to a very small size of about 0.3 km, compared to 4.3 km for the case 
$m_{\rm DM}=1$ GeV. 
Also, the density of the DM core is much higher than that of the NM for 
$m_{\rm DM}=7$ GeV.

%Also, we study the sensitivity of the radius of a DANS with the same 
%$M_{DM}$ but with different $m_{DM}$. In Fig. \ref{fig:radius_ratio_mass}
%we show the relative changes in radius against $M_{DM}$ for different 
%DANS models using $m_{DM}$ as a parameter.
%We find that the curves lie in a narrow band for stellar models with the 
%same $M_{DM}$. Similar result is found when studying 
%the relative change in the moment of inertia. Hence
%this again verifies our argument in the previous section studying
%the linear response, that
%the relative changes of the radius and the moment of inertia depend on 
%$M_{DM}$ and the stiffness of the EOS only in the linear regime. 

\begin{figure}
\centering
\includegraphics[width=8cm, height=6cm]{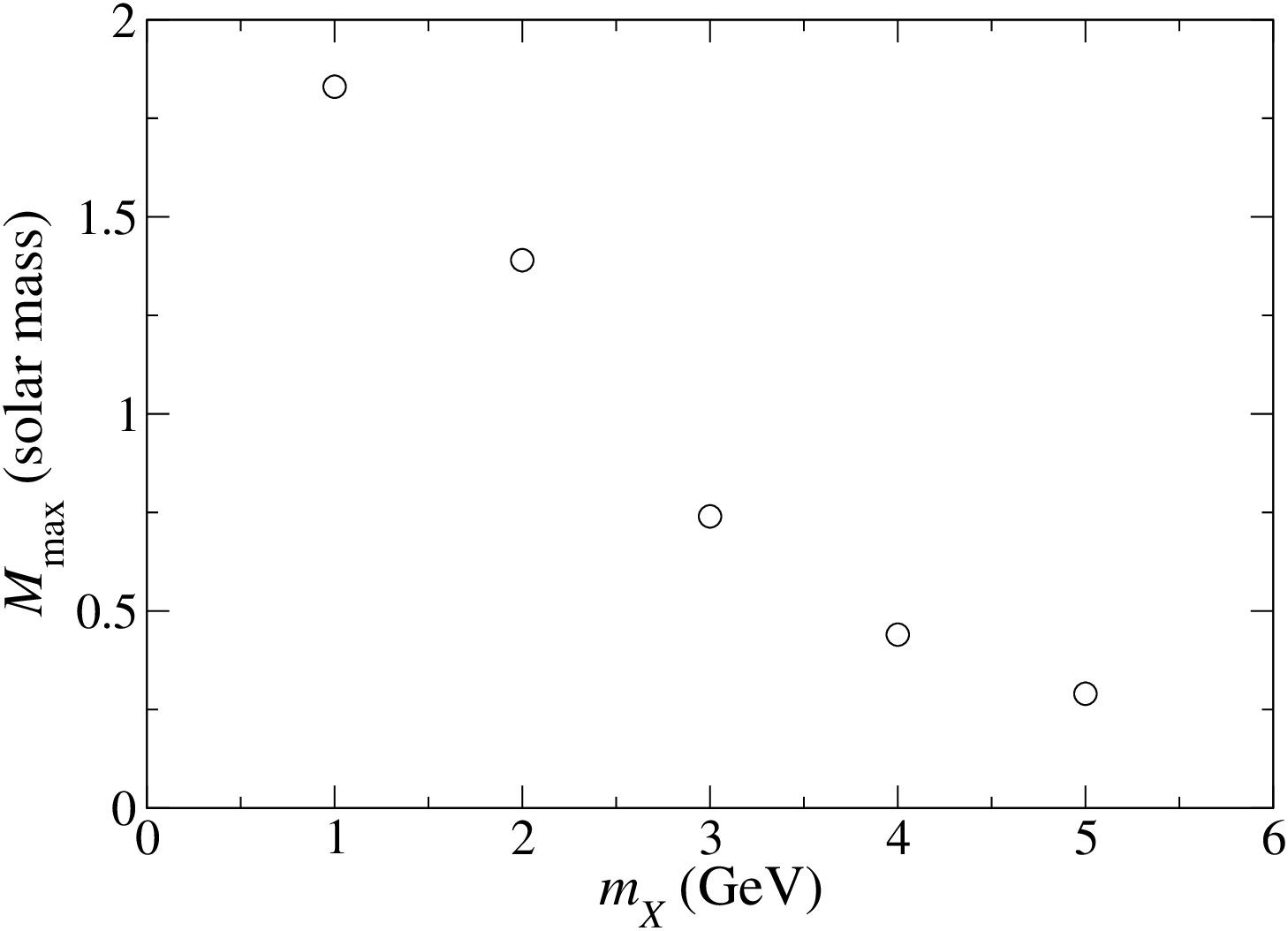}
\caption{Maximum stable mass $M_{\rm max}$ is plotted against the DM particle
mass $m_X$ for a fixed amount of DM specified by $\epsilon=0.1$. }
\label{fig:Mmax_eps0.1}
\end{figure}

%\begin{figure}
%\centering
%\includegraphics[width=8cm, height=6cm]{mode_all_one_fluid.eps}
%\caption{Lowest six eigenfunctions for the radial oscillation
%of a two-fluid star with the same surface for the polytropic 
%EOS for both NM and DM (see the text for the details).
%The numbers stand for the order of eigenfunction.
%We normalize the functions by setting the maximum displacement
%in the first eigenfunction to be unity.   
%$M = 1.183 M_{\odot}$, $M_{NM} = 0.928$ and $M_{DM} = 0.371$.}
%\label{fig:mode_all_samesurface}
%\end{figure}

Finally, in Fig. \ref{fig:Mmax_eps0.1} we plot 
the maximum stable mass for DANS models with $\epsilon = 0.1$
but with $m_{\rm DM}$ ranging from 1 to 5 GeV.
The maximum stable mass decreases with $m_{\rm DM}$. 
The reason is as follows:
the NM and DM are assumed to be noninteracting (except
through gravity), and the DM core is supported only by its
own degenerate pressure. It is well known that the maximum
mass limit for a self-gravitating Fermi gas decreases as the particle 
mass increases.
Hence, the onset of the collapse of a degenerate DM core is responsible for the
dependence of $M_{\rm max}$ on $m_{\rm DM}$ as seen in 
Fig. \ref{fig:Mmax_eps0.1}.
Furthermore, the allowed mass fraction of DM inside a stable DANS decreases 
significantly as $m_{\rm DM}$ increases. For example, stable DANS models with 
$M_{\rm DM} \approx 0.1 M_\odot$ can only be formed by DM particles of mass 
less than about 3 GeV.

\begin{figure}
\centering
\subfigure[]{
\includegraphics[width=8cm, height=6cm]{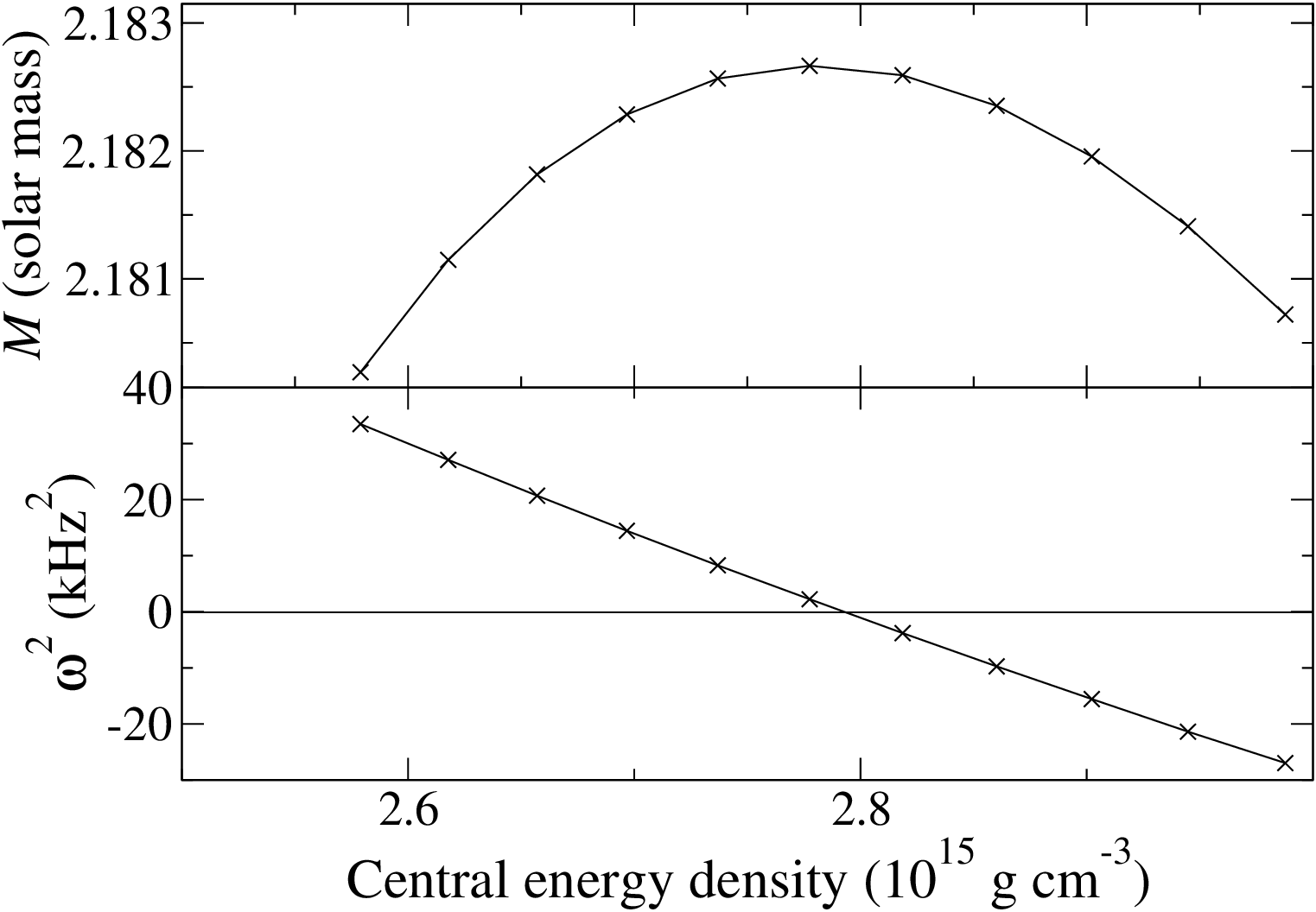}
\label{fig:eigenvalue1n_onefluid_NM}
}
\\
abc
\\
\subfigure{
\includegraphics[width=8cm, height=7cm]{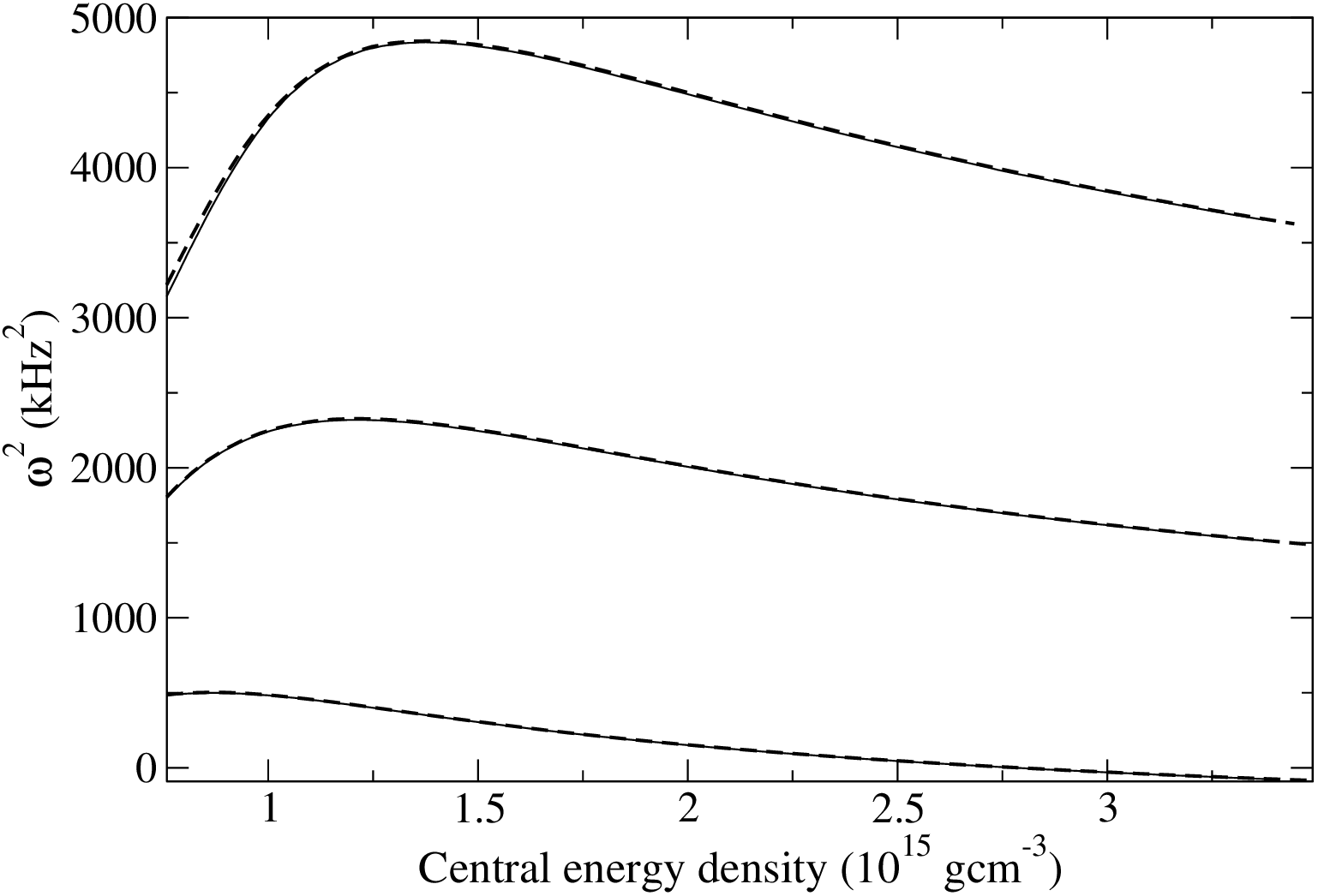}
\label{fig:eigenvalue_1fluid_compare}
}
\caption{Upper plot (a): Total mass (upper panel) and fundamental mode frequency 
squared (lower panel) are plotted against the central energy density for 
ordinary NS modeled by the APR EOS. 
Lower plot (b): Frequency squared for the first three modes are
plotted against the central energy density for ordinary NS modeled by the
APR EOS. The dashed lines are obtained by the two-fluid code in the one-fluid
limit. The solid lines are obtained by a
one-fluid code (see text). }
\end{figure}

%As a summary, we find that $R_{DM}$ changes inversely as $m_{DM}$
%for a DANS with given $M$ and $M_{DM}$
%and the relative changes of radius and moment of 
%inertia are not sensitive to $m_{DM}$ for a
%given DANS with the same $M_{NM}$. 
%Also, the minimum radius of a DANS
%with a given $M$ increases with $m_{DM}$.
%Therefore, to construct a 1.4 $M_{\odot}$ and 
%10 km radius DANS, we can use $m_{DM}$  
%not larger than 4 GeV. Also, the higher $m_{DM}$ is,
%the smaller the DM core becomes for the same $M_{DM}$. The 
%relative changes in radius and moment of inertia are weakly 
%related to $m_{DM}$. 

%\begin{figure}
%\centering
%\includegraphics[width=8cm, height=7cm]{fig8.eps}
%\caption{Frequency squared for the first three modes are 
%plotted against the central energy density for ordinary NS modeled by the 
%APR EOS. The dashed lines are obtained by the two-fluid code in the one-fluid
%limit (by setting the central number density of DM to be eight orders of 
%magnitude smaller than that of NM). The solid lines are obtained by a 
%one-fluid code (see text).  }
%\label{fig:eigenvalue_1fluid_compare}
%\end{figure}

%%%%%%%%%%%%%%%%%%%%%%%%%%%%%%%%%%%%%%%%%%%%%%%%%%%%%%%%%%%%%%%%%%%
\section{Radial Oscillations of DANS}
\label{sec:radial_results}

%%%%%%%%%%%%%%%%%%%%%%%%%%%%%%%%%%%%%%%%%%%%%%%%%%%%%%%%%%%%%%%%%%%

\begin{figure}
\centering
\subfigure[]{
\includegraphics[width=8cm, height=7cm]{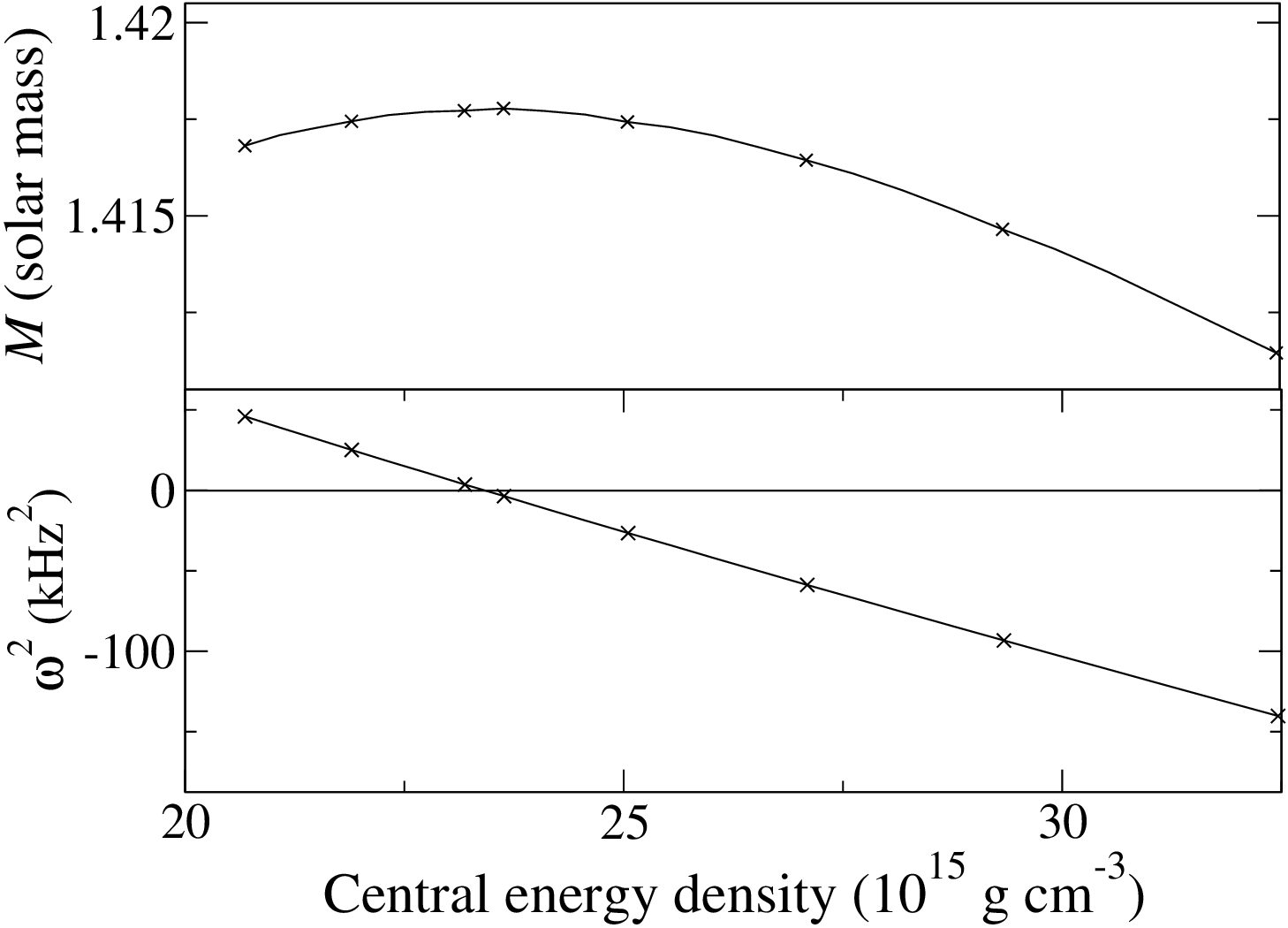}
\label{fig:eigenvalue_nc_02}
} 
\\ 
abc
\\
\subfigure{
\includegraphics[width=8cm, height=6cm]{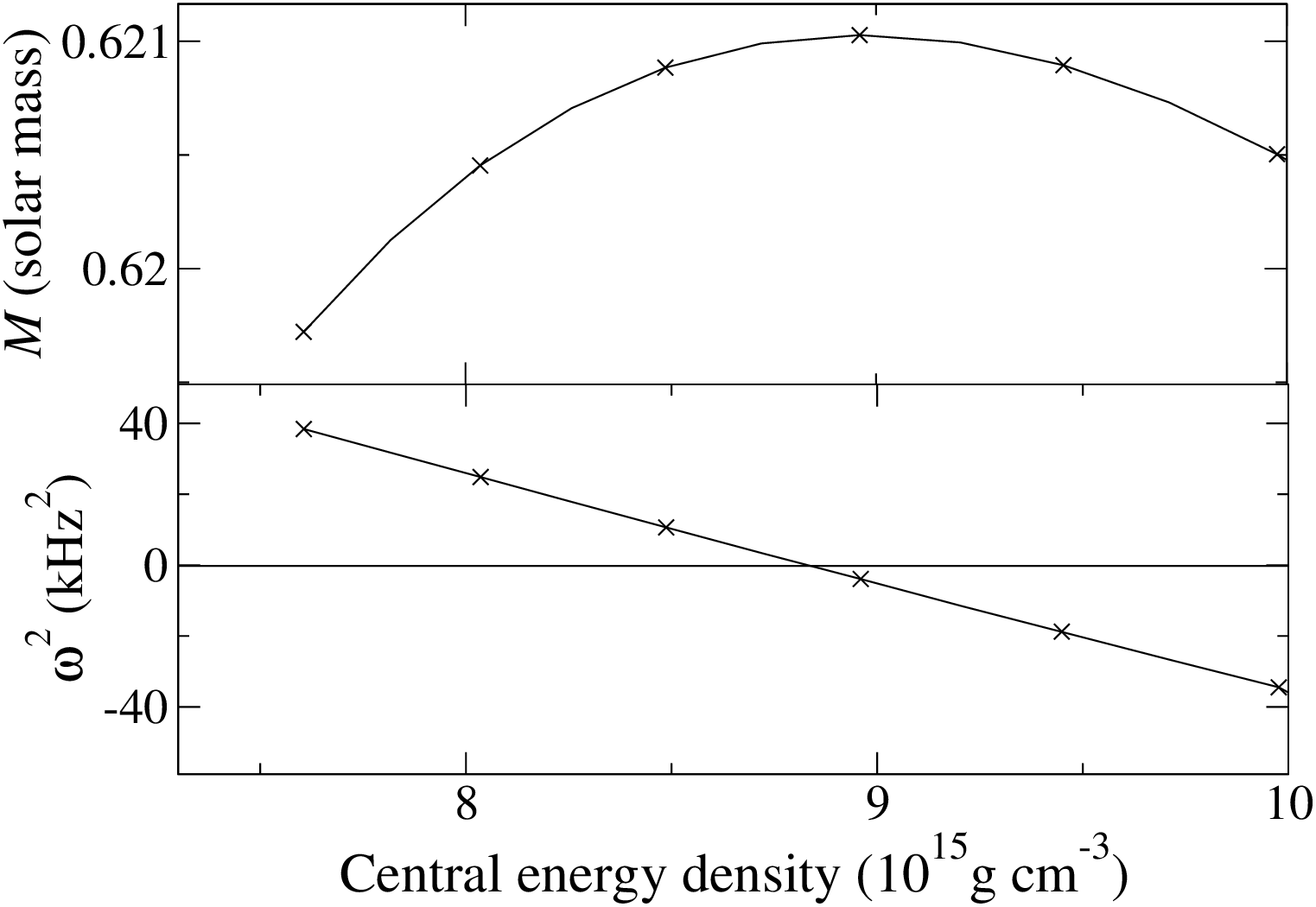}
\label{fig:eigenvalue_nc_08}
}
\caption{Upper plot (a):
Total mass and fundamental mode frequency squared
are plotted against the central energy density for $\epsilon=0.2$.
Lower plot (b):
Same as Fig. \ref{fig:eigenvalue_nc_02}, but for $\epsilon = 0.8$.
}
\end{figure}

In this section we study the radial oscillation eigenfrequencies and 
eigenfunctions of DANS. We show that all DANS with central energy 
density less than that of the maximum mass configuration, regardless of the
mass fraction $\epsilon$ of DM, are stable. 
We also study the effect of $\epsilon$ and DM particle mass $m_{\rm DM}$ on
the radial oscillation modes. 
In this section, we use the APR EOS to model the NM. Unless otherwise 
noted, the DM particle mass $m_{\rm DM}$ is chosen to be 1 GeV.

%%%%%%%%%%%%%%%%%%%%%%
\subsection{One-fluid limit}
%%%%%%%%%%%%%%%%%%%%%%

We first present 
some tests to check the validity of the numerical code.
We calculate the oscillation modes of ordinary NS modeled by the APR 
EOS in the one-fluid limit using our two-fluid code. In practice, this is 
achieved by setting the central density of DM to a sufficiently small number 
so that the star is essentially composed of NM only. 
In Fig.~\ref{fig:eigenvalue1n_onefluid_NM} we plot the total mass (upper 
panel) and fundamental mode frequency squared (lower panel) 
against the central energy density. 
We set the central number density of DM to be 8 orders of magnitude 
smaller than that of NM in the calculations. 
As expected from the study of ordinary NS, 
Fig.~\ref{fig:eigenvalue1n_onefluid_NM} shows that the fundamental mode
frequency passes through zero at the central energy density corresponding to 
the maximum mass configuration. The point $\omega^2=0$ marks the onset of 
instability. Beyond this critical density, the stars are unstable against 
radial perturbations. 

We compare in Fig.~\ref{fig:eigenvalue_1fluid_compare} the 
frequencies of the first three modes calculated separately by the 
two-fluid code (dashed lines) and a different code (solid lines) based on the 
standard one-fluid formulation \cite{Chandrasekhar:1}. 
We see that the two sets of mode frequencies agree very well.

\begin{figure}
\centering
\subfigure[]{
\includegraphics[width=8cm, height=6cm]{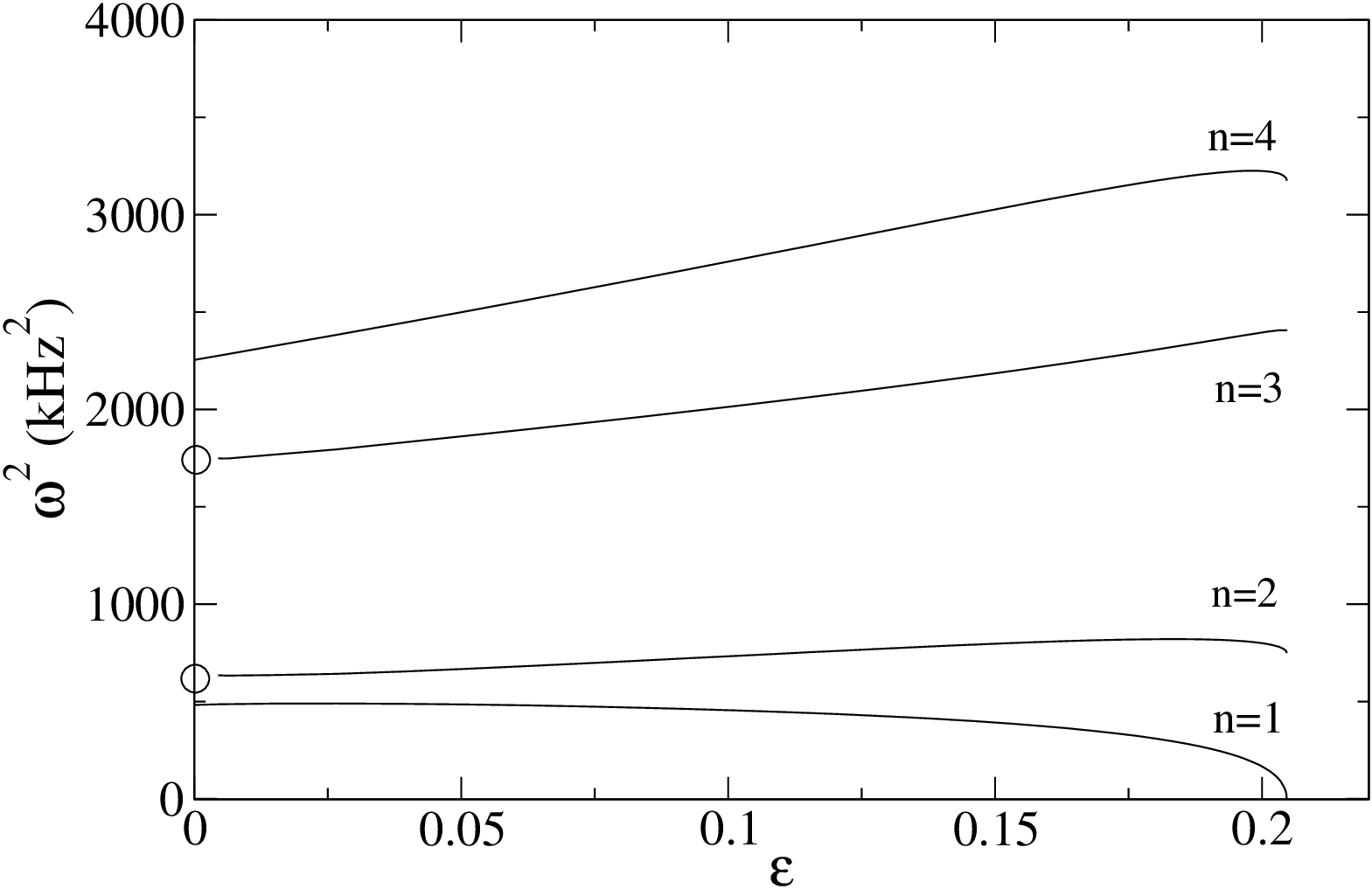}
\label{fig:eigenvalue_all_14_plot}
}
\\
abc
\\
\subfigure{
\includegraphics[width=8cm, height=6cm]{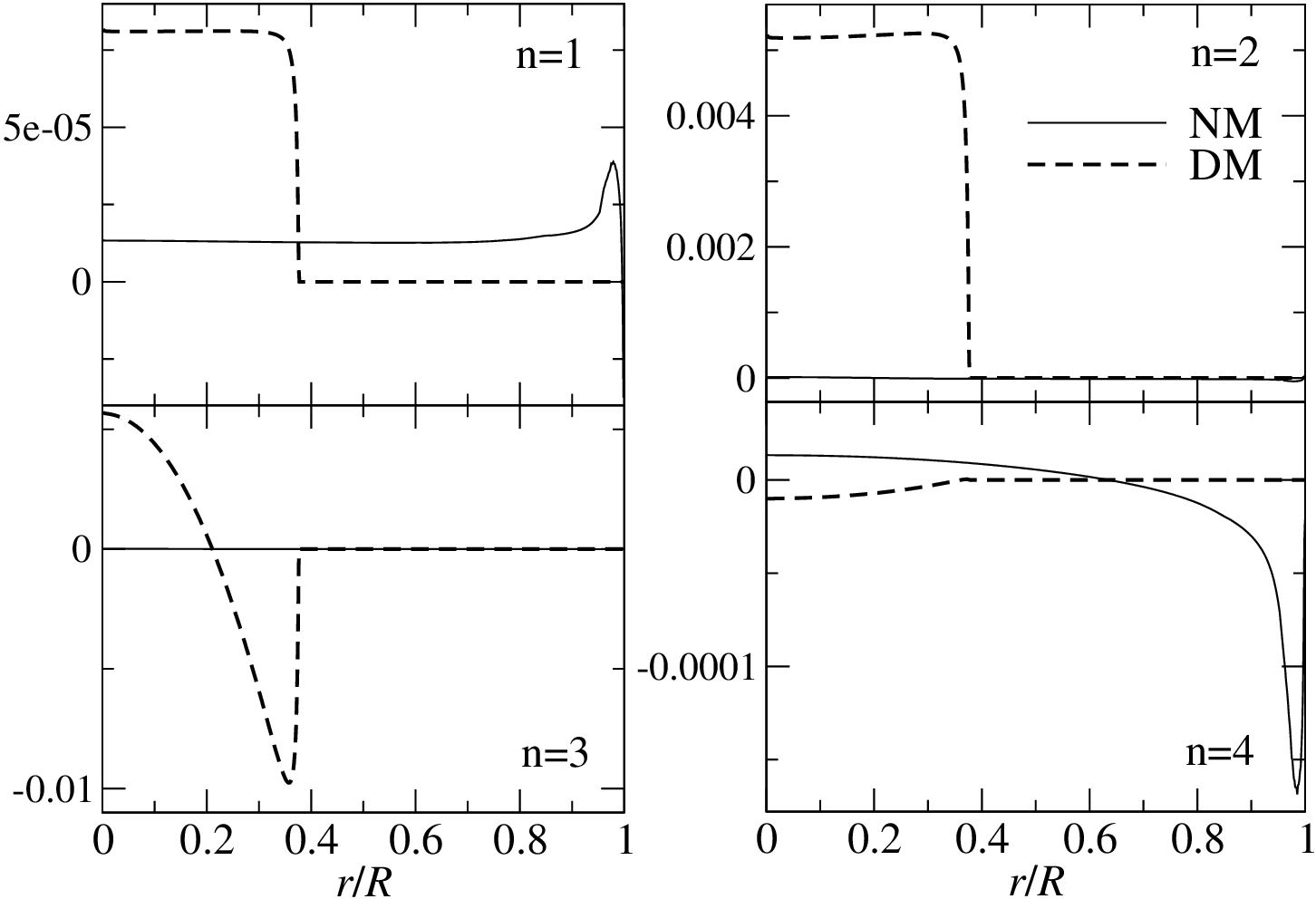}
\label{fig:eigenvalue_all_14_plot2}
}

\caption{Upper plot (a):
The frequency squared for the first four oscillation modes are plotted
against $\epsilon$. The total mass $M=1.4 M_\odot$ is fixed. 
The circles indicate the absence of the corresponding modes at the limit 
$\epsilon = 0$.
Lower plot (b):
The Lagrangian density variations of the first four modes
of a DANS with $\epsilon = 0.005$ and total mass $M = 1.4 M_{\odot}$.}
\end{figure}

%%%%%%%%%%%%%%%%%%%%%%%%%%%%%%%%%%%%%%%%%%%%%%%%%%%%%%%%%%%%%%%%%%
\subsection{Oscillation modes of DANS}
%%%%%%%%%%%%%%%%%%%%%%%%%%%%%%%%%%%%%%%%%%%%%%%%%%%%%%%%%%%%%%%%%

In Fig. \ref{fig:eigenvalue_nc_02}, we plot the total mass 
(upper panel) and the fundamental mode frequency squared (lower panel)
against the central energy density for $\epsilon = 0.2$. 
Similar to the ordinary NS (one-fluid) case, the mode frequency passes through
zero at the central energy density corresponding to the maximum mass 
configuration. The stars are unstable beyond this critical density. For 
DANS with central density lower than the critical density, they are stable 
against radial perturbations. 
In Fig. \ref{fig:eigenvalue_nc_08} we show the case $\epsilon=0.8$ for 
the DM dominated sequence.  
Our results confirm the stability of DANS. In particular, the new class of 
DM dominated compact stars with a NM core embedded in a ten-kilometer sized
DM halo are shown to be stable.

\begin{figure}
\centering
\includegraphics[width=8cm, height=6cm]{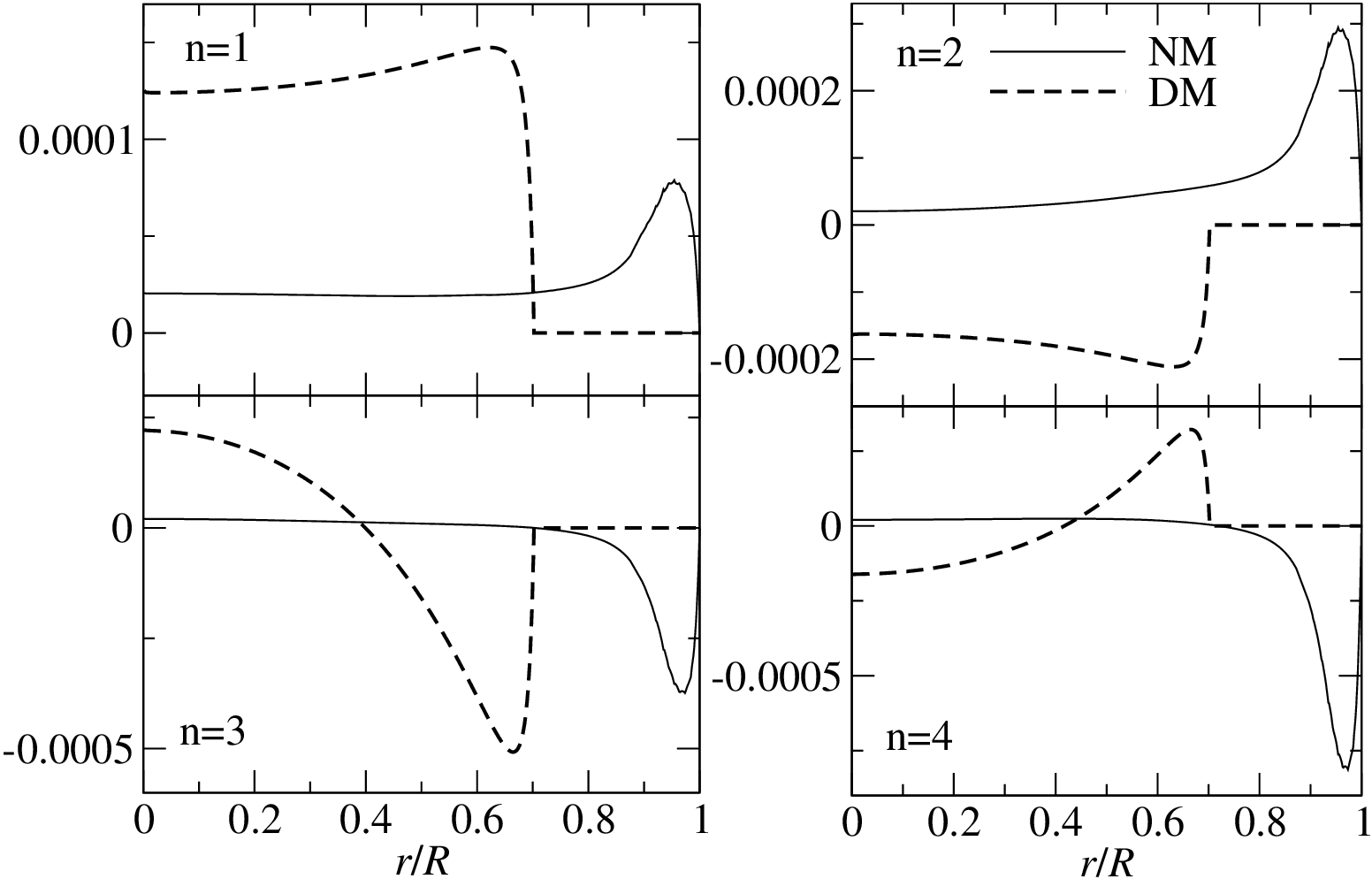}
\caption{The Lagrangian density variations of the first four modes for the 
ordinary DANS with $\epsilon=0.1$ at the point Y in 
Fig.~\ref{fig:MR_relation}.}
\label{fig:mode_all_04_03}
\end{figure}

In Fig. \ref{fig:eigenvalue_all_14_plot} we plot the first four mode 
frequency squared of DANS as a function of $\epsilon$. The total mass 
$M=1.4 M_\odot$ is fixed. 
We can see from the frequency of the fundamental mode ($n=1$) that increasing
$\epsilon$ (i.e., the mass fraction of DM) has the effect of decreasing the 
stability of the star. The fundamental mode frequency drops sharply to zero
for $\epsilon$ slightly above 0.2, hence indicating the onset of 
instability. 
We also see that the frequencies of the higher order modes ($n=2, 3, 4$) 
in general increase with $\epsilon$. 
It is interesting to notice that the second ($n=2$) and third ($n=3$)
modes are missing in the one-fluid limit when $\epsilon=0$. 
In Fig. \ref{fig:eigenvalue_all_14_plot2},
we show the Lagrangian variations of NM [Eq.~(\ref{eq:delta_n_over_n})] 
and DM [Eq.~(\ref{eq:delta_p_over_p})] for the first four modes of a star 
with $\epsilon = 0.005$. The solid (dashed) lines are the profiles of NM (DM). 
Note that the DM core extends to about $0.4R$, where $R$ is the radius of the
star. 

First let us consider the $n=1$ and $n=4$ modes. They have proper 
limits at $\epsilon=0$. 
For the fundamental mode ($n=1$), in the DM core where the two fluids
coexist, the density variations of NM and DM are in phase. For the $n=4$ mode,
the density variation of NM (DM) is larger (smaller) than zero in the DM core. 
Hence, the two fluids are counter-moving in this case. On the other hand, 
for the $n=2$ and $n=3$ modes, we see that the density variation of NM is 
much smaller than that of DM. These modes are dominated by DM motion in 
the DM core. 
They do not exist when there is no DM (i.e., $\epsilon=0$). 
However, they emerge even for a very small mass fraction of DM.

To further study the oscillation modes of DANS, we choose the two stellar
models at the point Y in Fig. \ref{fig:MR_relation}.  
The stars have the same mass $M=0.541 M_\odot$ and radius $R=9.68$ km, but
with different mass fraction of DM. The model with $\epsilon=0.1$ is an 
ordinary DANS. The model with $\epsilon=0.8$ is a DM dominated compact star 
with a small NM core embedded in a ten-kilometer sized DM halo. 
The Lagrangian variations of the first four oscillation modes for the model 
with $\epsilon=0.1$ are plotted in Fig.~\ref{fig:mode_all_04_03}. 
For the $n=1$ and $n=4$ modes, the general patterns of the modes are 
qualitatively the same as the case $\epsilon=0.005$ in 
Fig.~\ref{fig:eigenvalue_all_14_plot2}.  
The two fluids are in large part comoving (counter-moving) for the 
$n=1$ ($n=4$) mode. 
For the $n=2$ and $n=3$ modes, which are dominated by 
DM motion in the case $\epsilon=0.005$, we now see that there are
significant density variations of NM near the stellar surface. 
This can be understood by the fact that the two fluids are coupled through 
gravity. 
For a very small $\epsilon$, the motion of a small amount of DM basically
has no effect on the NM. Hence, the NM essentially decouples from the DM and 
does not move in the $n=2$ and $n=3$ DM dominated modes.  
However, when the mass fraction of DM is comparable to NM, the coupling 
between the two fluids becomes stronger and hence we can see a large density 
variation of NM. 

\begin{figure}
\centering
\includegraphics[width=8cm, height=6cm]{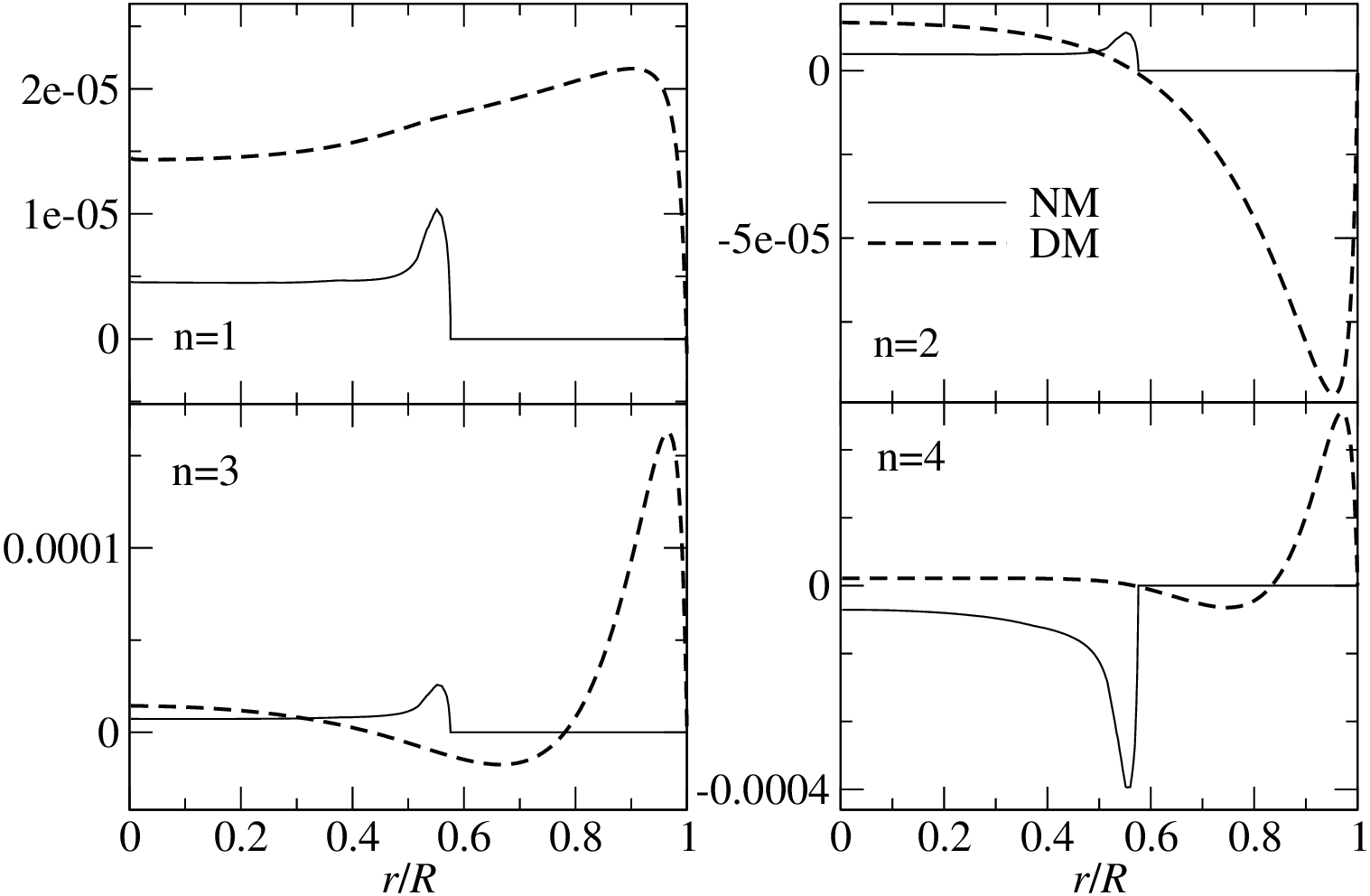}
\caption{
The Lagrangian density variations of the first four modes for the 
DM dominated compact star with $\epsilon=0.8$ at the point Y in 
Fig.~\ref{fig:MR_relation}.} 
\label{fig:mode_all_04_06}
\end{figure}

The Lagrangian variations of the first four oscillation modes for the 
DM dominated model with $\epsilon=0.8$ are plotted in 
Fig.~\ref{fig:mode_all_04_06}. We see that for the ``ordinary'' $n=1$ and 
$n=4$ modes the maximum density variations of the two fluids near their 
surfaces are comparable. 
Note that the surface of the NM core is at about $0.6R$, where the radius 
of the star $R$ is defined by the radius of the DM halo in this model. 
However, for the $n=2$ and $n=3$ DM dominated modes, the maximum
density variation of DM is much larger than that of NM.

%%%%%%%%%%%%%%%%%%%%%%%%%%%%%%%%%%%%%%%%%%%%%%%%%%%%%%%%%%%%%%%%%%%%%%%%%
\subsection{Effects of DM particle mass}
%%%%%%%%%%%%%%%%%%%%%%%%%%%%%%%%%%%%%%%%%%%%%%%%%%%%%%%%%%%%%%%%%%%%%%%%%

\begin{figure}
\centering
\includegraphics[width=8cm, height=6cm]{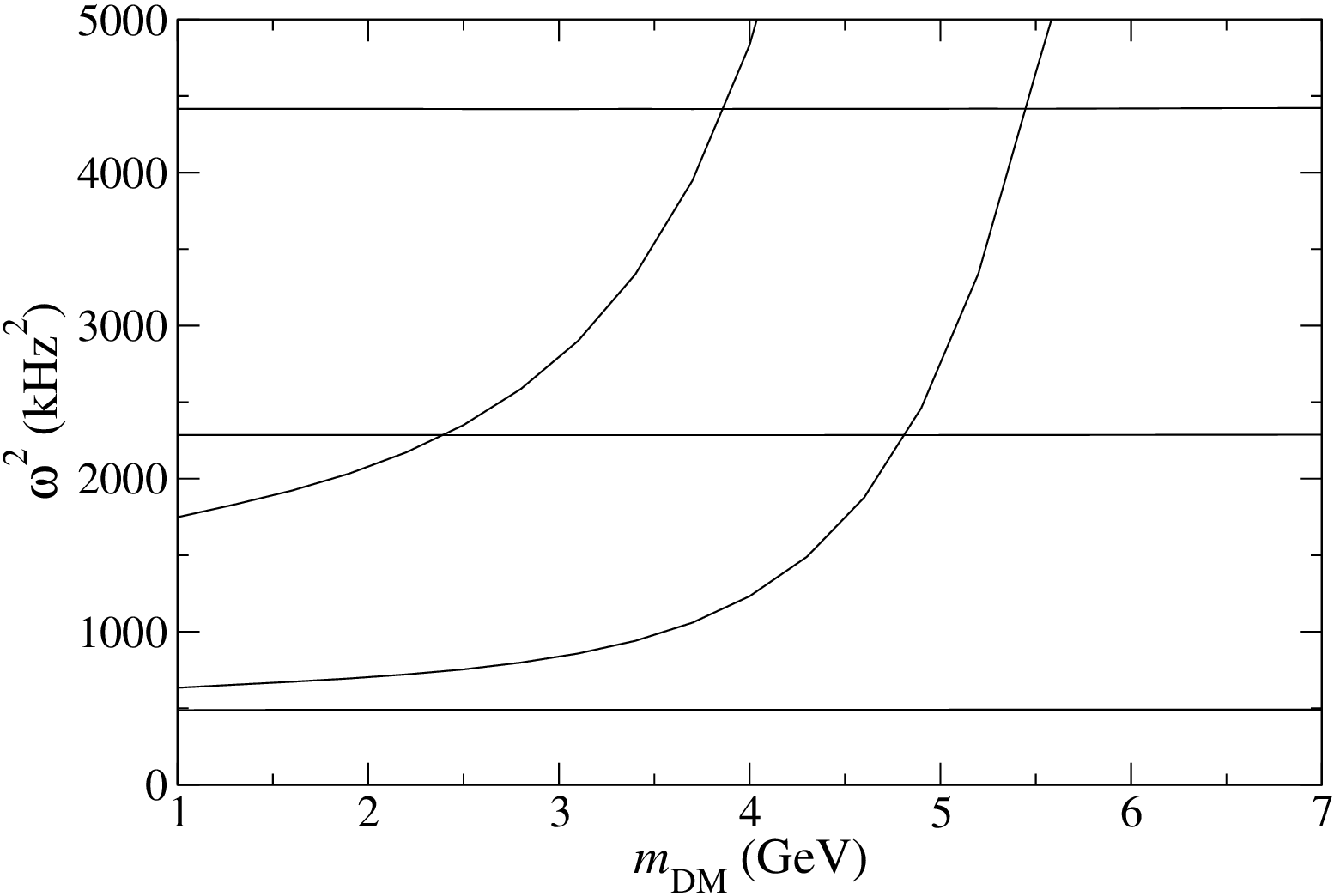}
\caption{
Frequency squared for the first five oscillation modes are plotted
against the DM particle mass $m_{\rm DM}$.
The total mass $M=1.4 M_\odot$ and DM core mass $M_{\rm DM}=0.01
M_\odot$ of the stars are fixed. }
\label{fig:eigenvalue_all_14_mass}
\end{figure}

To end this section, we study the effects of DM particle mass on 
the mode frequencies. 
In Fig. \ref{fig:eigenvalue_all_14_mass} we show the frequencies of 
the first five modes of a DANS with $M = 1.4 M_{\odot}$ and 
$M_{DM} = 0.01 M_{\odot}$ as a function of DM particle mass $m_{\rm DM}$.
For $m_{\rm DM}=1$ GeV, the second and third modes correspond to the 
$n=2$ and $n=3$ DM dominated modes studied above. We see that the frequencies 
of these modes increases with $m_{\rm DM}$, while the other modes are
essentially independent of $m_{\rm DM}$. 
It is known that the radial oscillation mode frequency (squared) scales with 
the density of the star. Also, for a fixed DM core mass, the DM core becomes 
denser as $m_{\rm DM}$ increases (see Fig.~\ref{fig:density_profile_mass}). 
Hence, the frequencies of the DM dominated modes depend strongly 
on $m_{\rm DM}$.

It is also interesting to notice that, while the second and third modes 
are DM dominated modes in the case $m_{\rm DM}=1$ GeV, this is in general
not true for other $m_{\rm DM}$. For example, in the 
case $m_{\rm DM}=3$ GeV, it is the second and fourth modes that are 
DM dominated modes. For $m_{\rm DM}=5$ GeV, the DM dominated mode first 
appears only in the third mode.

%%%%%%%%%%%%%%%%%%%%%%%%%%%%%%%%%%%%%%%%%%%%%%%%%%%%%%%%%%%%%%
\section{Conclusions}
\label{sec:conclude} 
%%%%%%%%%%%%%%%%%%%%%%%%%%%%%%%%%%%%%%%%%%%%%%%%%%%%%%%%%%%%%%

In this paper, we have studied the equilibrium properties and radial 
oscillation modes of DANS using a general relativistic two-fluid formalism. 
We model the NM by realistic nuclear matter EOS. The DM particles are 
assumed to be non-self-annihilating and described by an ideal degenerate
Fermi gas. Our results suggest that the structure of these stars depends 
strongly on the DM fluid. In particular, we found a new class of compact 
stars which are dominated by DM. These stars in general have a small NM core 
with radius a few km embedded in a larger ten-kilometer-sized DM halo. 
Since only the NM core can emit thermal radiation, the detection of a compact 
star with a thermally radiating surface of such a small size could 
provide a strong evidence for the existence of DANS. Furthermore, these DM 
dominated stars also have rather different mass-radius relations and 
scaled moment of inertia compared to ordinary NS without DM. 
We have also studied how the radius $R$ and moment of inertia $I$ 
of a star with fixed NM baryonic mass $M_{\rm NM}$ change as the DM core 
mass $M_{\rm DM}$ increases. In the small $M_{\rm DM}$ limit, 
we see that $R$ and $I$ decrease linearly as $M_{\rm DM}$ increases. The 
slopes of the linear variations depend essentially only on the NM EOS, but 
not on the value of $M_{\rm NM}$.

We have performed a radial perturbation analysis and studied the oscillation
modes of DANS in general. The stability of DANS is shown explicitly by 
calculating the frequency of the fundamental mode. For a sequence of stars
with a fixed DM mass fraction, we see that the fundamental mode frequency 
passes through zero at the central energy density corresponding to the 
maximum mass configuration. Similar to the analysis of ordinary NS, this point 
marks the onset of instability.
 
Besides the fundamental mode, we have also studied the first few higher 
order oscillation modes. 
We see that DANS in general have two classes of oscillation modes. The first 
class of modes has proper limit when the DM mass fraction tends to 
zero, namely these modes reduce to the same set of modes for ordinary NS 
without DM. 
On the other hand, the second class of modes is due mainly to DM. 
In the limit of a small DM mass fraction, these modes are characterized purely 
by the oscillations of DM. The NM fluid is essentially at rest. 
In the intermediate case where the mass fractions of NM and DM are comparable, 
the NM fluid oscillates with the DM fluid due to their coupling through 
gravity. 
On the other hand, the amplitude of DM oscillations is much larger than 
that of NM in the case of DM dominated stars. 
We also see that the frequencies of these oscillation modes depend strongly 
on the DM particle mass. 

Finally, it should be pointed out that the formation mechanism of DANS is not clear. 
However, our main focus in this work is to study the properties of these
theoretical objects, if they exist. The formation process of these compact 
dark-matter compact objects cannot yet be modeled in current $N$-body 
simulations, which mainly focus on the structure formation in galactic or 
cosmological scales.

%For future studies, the effect of
%DM core on the stellar formation and
%evolution is of great interest.
%Several questions are of interest:
%what are the effects of the DM core on the
%stellar lifetime and its final fate?
%Can the DM core lower the current mass limits for
%main-sequence star, white dwarf
%and NS formation? What is the role
%of the DM core in the supernova process?
%Our study can be further extended
%to bosonic DM or even self-annihilating
%DM. These studies may lead to
%tighter constraints on   
%DM particle properties.

\acknowledgments
We thank D.-L. Cheng for useful discussions. This work is partially supported 
by a grant from the Research Grant Council of the Hong Kong Special 
Administrative Region, China (Project No. 400910).

%\newpage

%%%%%%%%%%%%%%%%%%%%%%%%%%%%%%%%%%%%%% 
%% reference 
%%%%%%%%%%%%%%%%%%%%%%%%%%%%%%%%%%%%%% 
\bibliographystyle{prsty}

\begin{thebibliography}{1}

\bibitem[\protect\citeauthoryear{M. Roos}{2010}]{Roos:1}
M. Roos, arXiv:1001.0316v2

\bibitem[\protect\citeauthoryear{Y. Sofue et al.}{2009}]{Sofue:1}
Y. Sofue, M. Honma and T. Omodaka, PASJ {\bf61}, 227 (2009).

\bibitem[\protect\citeauthoryear{R. Catana et al.}{2010}]{Catena:1}
R. Catena and P. Ullio, JCAP {\bf08}, 004 (2010).

\bibitem[\protect\citeauthoryear{M. Weber et al.}{2010}]{Weber:1}
M. Weber, and W. de Boer, A \& A {\bf509}, A25 (2010).

\bibitem[\protect\citeauthoryear{R.H. Miller et al.}{1970}]{Miller:1}
R. H. Miller, K. H. Prendergast and W. J. Quirk,  ApJ {\bf161}, 903 (1970).

\bibitem[\protect\citeauthoryear{F. Hohl}{1971}]{Hohl:1}
F. Hohl, ApJ {\bf168}, 343 (1971).

\bibitem[\protect\citeauthoryear{J.P. Ostriker et al.}{1973}]{Ostriker:1}
J. P. Ostriker and P. J. E. Peebles, ApJ {\bf186}, 467 (1973).

%\bibitem[\protect\citeauthoryear{G. Efstathiou et al.}{1984}]{Efstathiou:1}
%G. Efstathiou and J. R. Bond, Phil. Trans. Roy. Soc. London, Series A. Math. Phys. Sci, {\bf320}, 585 (1986).

\bibitem[\protect\citeauthoryear{V. Springel et al.}{2005}]{Springel2005}
V. Springel {\it et al.}, Nature {\bf435}, 629 (2005).

\bibitem[\protect\citeauthoryear{R. Massey et al.}{2007}]{Massey:1}
R. Massey {\it et al.}, Nature {\bf445}, 286 (2007).

%\bibitem[\protect\citeauthoryear{J.L. Feng et al.}{2010}]{Feng:2}  
%J. L. Feng, Ann. Rev. of Astron. and Astrophys., {\bf48}, 495 (2010).

\bibitem[\protect\citeauthoryear{R. Bernabei et al.}{2008}]{DAMA2008}
R. Bernabei {\it et al.} (DAMA Collaboration), Eur. Phys. J. C {\bf 56}, 333 (2008).

\bibitem[\protect\citeauthoryear{C.~E. Aalseth et al.}{2011}]{CoGeNT2011}
C.~E. Aalseth {\it et al.} (CoGeNT Collaboration)Phys. Rev. Lett. {\bf 106},
131301 (2011).

\bibitem[\protect\citeauthoryear{Z. Ahmed et al.}{2011}]{CDMS2011}
Z. Ahmed {\it et al.} (CDMS Collaboration), Phys. Rev. Lett. {\bf 106}, 131302
(2011).

\bibitem[\protect\citeauthoryear{E. Aprile et al.}{2010}]{XENON2010} 
E. Aprile {\it et al.} (XENON100 Collaboration), Phys. Rev. Lett. {\bf 105},
131302 (2010).

\bibitem[\protect\citeauthoryear{J.~L. Feng et al.}{2011}]{Feng2011}
J.~L. Feng {\it et al.}, Phys. Lett. B {\bf 703}, 124 (2011).

\bibitem[\protect\citeauthoryear{M.~T. Frandsen et al.}{2011}]{Frandsen2011}
M.~T. Frandsen {\it et al.}, Phys. Rev. D {\bf 84}, 041301 (2011).

\bibitem[\protect\citeauthoryear{D. Spolyar et al.}{2008}]{Spolyar2008}
D. Spolyar, K. Freese and P. Gondolo, Phys. Rev. Lett. {\bf 100}, 051101 (2008).

\bibitem[\protect\citeauthoryear{D. Spolyar et al.}{2009}]{Spolyar2009}
D. Spolyar, P. Bodenheimer, K. Freese and P. Gondolo, Astrophys. J. {\bf 705}, 1031 (2009).

\bibitem[\protect\citeauthoryear{E. Ripamonti et al.}{2010}]{Ripamonti2010}
E. Ripamonti, F. Iocco, A. Ferrara, R. Schneider, A. Bressan and P. Marigo, Mon. Not. R. Astron. Soc. {\bf 406}, 2605 (2010).

\bibitem[\protect\citeauthoryear{S. Hirano et al.}{2011}]{Hirano2011}
S. Hirano, H. Umeda and N. Yoshida, ApJ {\bf 736}, 58 (2011).

\bibitem[\protect\citeauthoryear{J. Casanellas et al.}{2009}]{Casanellas2009}
J. Casanellas and I. Lopes, ApJ {\bf 705}, 135 (2009).

\bibitem[\protect\citeauthoryear{J. Casanellas et al.}{2011}]{Casanellas2011}
J. Casanellas and I. Lopes, ApJ {\bf 733}, L51 (2011).

\bibitem[\protect\citeauthoryear{I. Lopes et al.}{2011}]{Lopes2011}
I. Lopes, J. Casanellas and D. Eugenio, Phys. Rev. D {\bf 83}, 063521 (2011).

\bibitem[\protect\citeauthoryear{A. H. G. Peter}{2009}]{Peter2009}
A. H. G. Peter, Phys. Rev. D {\bf 79}, 103531 (2009).

\bibitem[\protect\citeauthoryear{L. Iorio et al.}{2010}]{Iorio2010a}
L. Iorio, JCAP {\bf 1005}, 018 (2010).

\bibitem[\protect\citeauthoryear{L. Iorio et al.}{2010}]{Iorio2010b} 
L. Iorio, JCAP {\bf 1011}, 046 (2010).

\bibitem[\protect\citeauthoryear{M.~T. Frandsen et al.}{2010}]{Frandsen2010}  
M.~T. Frandsen and S. Sarkar, Phys. Rev. Lett. {\bf 105}, 011301 (2010).

\bibitem[\protect\citeauthoryear{D.~T. Cumberbatch et al.}{2010}]{Cumberbatch2010}
D.~T. Cumberbatch {\it et al.}, Phys. Rev. D {\bf 82}, 103503 (2010).

\bibitem[\protect\citeauthoryear{M. Taoso et al.}{2010}]{Taoso2010}
M. Taoso {\it et al.}, Phys. Rev. D {\bf 82}, 083509 (2010).

\bibitem[\protect\citeauthoryear{I. Goldman et al.}{1989}]{Goldman1989}
I. Goldman and S. Nussinov, Phys. Rev. D {\bf 40}, 3221 (1989).

\bibitem[\protect\citeauthoryear{G. Bertone et al.}{2007}]{Bertone:1}
G. Bertone and M. Fairbairn, Phys. Rev. D {\bf77}, 043515 (2008).

\bibitem[\protect\citeauthoryear{M. McCullough et al.}{2010}]{McCullough2010}
M. McCullough and M. Fairbairn, Phys. Rev. D {\bf81}, 083520 (2010).

\bibitem[\protect\citeauthoryear{C. Kouvaris}{2008}]{Kouvaris2008}
C. Kouvaris, Phys. Rev. D {\bf 77}, 023006 (2008).

%\bibitem[\protect\citeauthoryear{Kholopov et al.}{2008}]{Kouvaris:2}
%M. Y. Kholopov and C. Kouvaris, Phys. Rev. D {\bf77}, 065002 (2008). 

\bibitem[\protect\citeauthoryear{Kholopov et al.}{2008}]{Kouvaris:2}
C. Kouvaris and P. Tinyakov, Phys. Rev. D {\bf82}, 063531 (2010).

\bibitem[\protect\citeauthoryear{A. Lavallaz et al.}{2010}]{Lavallaz2010}
A. de Lavallaz and M. Fairbairn, Phys. Rev. D {\bf81}, 123521 (2010).

\bibitem[\protect\citeauthoryear{D.E. Kaplan et al.}{2009}]{Kaplan:1}
D. E. Kaplan, M. A. Luty and K. M. Zurek, Phys. Rev. D {\bf79}, 115016 (2009).

\bibitem[\protect\citeauthoryear{F. Saudin et al.}{2009}]{Saudin2009}
F. Saudin and P. Ciarcelluti, Astropart. Phys. {\bf32}, 5, 278 (2009). 

\bibitem[\protect\citeauthoryear{P. Ciarcelluti et al.}{2011}]{Ciarcelluti2011}
P. Ciarcelluti and F. Sandin, Phy. Lett. B {\bf695}, 19 (2011).

\bibitem[\protect\citeauthoryear{S.-C. Leung et al.}{2011}]{Leung2011}
S.-C. Leung, M.-C. Chu and L.-M. Lin, Phys. Rev. D {\bf 84}, 107301 (2011).

\bibitem[\protect\citeauthoryear{B. Carter}{1989}]{Carter1989}
B. Carter, in {\it Relativistic Fluid Dynamics (Noto, 1987)},
edited by A. Anile and M. Choquet-Bruhat, Lecture Notes in Mathematics
Vol. 1385 (Springer-Verlag, Heidelberg, Germany, 1989), pp. 1-64.

\bibitem[\protect\citeauthoryear{G.~L. Comer et al.}{1999}]{Comer1999}
G.~L. Comer, D. Langlois and L.~M. Lin, Phys. Rev. D {\bf60}, 104025 (1999).

\bibitem[\protect\citeauthoryear{N. Andersson et al.}{2001}]{Andersson2001}
N. Andersson, G.~L. Comer, Class. Quantum Grav. {\bf18}, 969 (2001).

\bibitem[\protect\citeauthoryear{U. Lee}{1995}]{Lee:1}
N. Andersson, G. L. Comer and D. Langlois, Phys. Rev. D {\bf66}, 104002 (2002).

%\bibitem[\protect\citeauthoryear{G.L. Comer et al.}{1999}]{Lin:1}
%G. L. Comer, D. Langlois and L.-M. Lin, Phys. Rev. D, {\bf60}, 104025 (1999).

\bibitem[\protect\citeauthoryear{R. Prix et al.}{2001}]{Prix:1}
R. Prix, J. Novak and G. L. Comer, Phys. Rev. D {\bf71}, 043005 (2005).

\bibitem[\protect\citeauthoryear{A. Akmal et al.}{1998}]{Akmal:1}
A. Akmal, V. R. Pandharipande and D. G. Ravenhall, Phys. Rev. C {\bf 58}, 1804 (1998).

\bibitem[\protect\citeauthoryear{M. Baldo et al.}{1997}]{Baldo1997}   
M. Baldo, I. Bombaci and G. F. Burgio, A \& A {\bf 328}, 274 (1997).

\bibitem[\protect\citeauthoryear{J.B. Hartle et al.}{1974}]{Hartle:1}
J. B. Hartle, ApJ {\bf 150}, 1005 (1967).

\bibitem[\protect\citeauthoryear{M. Bejger et al.}{2002}]{Haensel:1}
M. Bejger and P. Haensel, A\&A {\bf396}, 917-921 (2002).

\bibitem[\protect\citeauthoryear{S. Chandrasekhar}{1964}]{Chandrasekhar:1}
S. Chandrasekhar, ApJ {\bf140}, 417 (1964).

%\bibitem[\protect\citeauthoryear{G. Chanmugam}{1977}]{Chanmugam:1}
%G. Chanmugam, ApJ {\bf217}, 799 (1977).

%\bibitem[\protect\citeauthoryear{E. N. Glass et al.}{1983}]{Glass1983}   
%E. N. Glass and L. Lindblom, ApJ {\bf53}, 93 (1983).

%\bibitem[\protect\citeauthoryear{H.M. Vath et al.}{1992}]{Vath1992}
%H. M. Vath and G. Chanmugam, Astron. Astrophys. {\bf 260}, 250 (1992).

%\bibitem[\protect\citeauthoryear{K.D. Kokkotas et al.}{2001}]{Kokkotas:3}
%K. D. Kokkotas and J. Ruoff, Astron.Astrophys. {\bf366}, 565 (2001).

%\bibitem[\protect\citeauthoryear{D. Gondek et al.}{1999}]{Gondek1999}
%D. Gondek and J. L. Zdunik, A \& A {\bf344}, 117 (1999).

%\bibitem[\protect\citeauthoryear{E. Gourgoulhon et al.}{1995}]{Gourgoulhon1995}
%E. Gourgoulhon, P Haensel and D. Gondek, A \& A {\bf294}, 747 (1995).

%\bibitem[\protect\citeauthoryear{D. Gondek et al.}{1997}]{Gondek1997}
%D. Gondek, P. Haensel and J. L. Zdunik, A \& A {\bf325}, 217 (1997).

%\bibitem[\protect\citeauthoryear{W. Laner et al.}{1969}]{Laner:1}
%W. D. Langer and A. G. W. Cameron, Astrophys. and Space Sci. {\bf 5}, 213 (1969).

%\bibitem[\protect\citeauthoryear{R. F. Sawyer}{1980}]{Sawyer1980}
%R. F. Sawyer, ApJ {\bf 237}, 187 (1980).

%\bibitem[\protect\citeauthoryear{C. Cutler}{1990}]{Cutler1990}
%C. Cutler, L. Lindblom and R. J. Splinter, ApJ {\bf 363}, 603 (1990).

%\bibitem[\protect\citeauthoryear{G. Narain et al.}{2006}]{Narain2006}
%G. Narain, J. Schaffner-Bielich and I. N. Mishustin, Phys. Rev. D. {\bf 74}, 063003 (2006).

%\bibitem[\protect\citeauthoryear{G.L. Comer et al.}{2001}]{Andersson2001}
%N. Andersson and G. L. Comer, Class. Quantum Grav. {\bf18}, 969-1002 (2001).

%\bibitem[\protect\citeauthoryear{R. Bernabei et al.}{2010}]{Bernabei:1}
%R. Bernabei {\it et al.}, 2010, Eur. Phys. J., {\bf C67}, 39-49

%\bibitem[\protect\citeauthoryear{C. E. Aalseth et al.}{2010}]{Aalseth:1}
%C. E. Aalseth {\it et al} (the CoGeNT Collaboration), 2010, arXiv:1002.4703

%\bibitem[\protect\citeauthoryear{Petriello et al.}{2008}]{Petriello:1}
%F. Petriello and K. M. Zurek, 2008, JHEP, {\bf0809}, 047

%\bibitem[\protect\citeauthoryear{P. Gondolo et al.}{2009}]{Gondolo:1}
%C. Savage, K. Freese, P. Gondolo and D. Spolyar, 2009, JCAP, {\bf0909}, 036

%\bibitem[\protect\citeauthoryear{P. Gondolo et al.}{2009}]{Gondolo:2}
%C. Savage, K. Freese, P. Gondolo and D. Spolyar, 2009, JCAP, {\bf0904}, 010

%\bibitem[\protect\citeauthoryear{M. Taoso et al.}{2010}]{Taoso:1}
%M. Taoso {\it et al}, 2010, arXiv:1005.5711v2

%\bibitem[\protect\citeauthoryear{D.T. Cumberbatch et al.}{2010}]{Cumberbatch:1}
%D. T. Cumberbatch {\it et al}, 2010, Phys. Rev. D., {\bf 82}, 103503

%\bibitem[\protect\citeauthoryear{D. Hooper et al.}{2010}]{Hooper:1}
%D. Hooper {\it et al.}, 2010, Phys. Rev. D, {\bf82}, 123509

%\bibitem[\protect\citeauthoryear{S. L. Shapiro et al.}{1983}]{Shapiro:1}
%S. L. Shapiro and S. A. Teukolsky, Black Holes, White Dwarfs and Neutron Stars, 1983 (New York: Wiley-Interscience)

%\bibitem[\protect\citeauthoryear{P.B. Demorest et al.}{2010}]{Demorest:1}
%P. B. Demorest {\it et al}, nature {\bf467}, 1081 (2010).

%\bibitem[\protect\citeauthoryear{Alcock et al.}{1984}]{Alcock:1}
%C. Alcock, E. Farhi and A. Olinto, ApJ {\bf310}, 261 (1986).

%\bibitem[\protect\citeauthoryear{F. Ozel et al.}{2009}]{Ozel:1}
%F. Ozel, T. Guver and D. Psaltis, ApJ {\bf693}, 1775 (2009).

%\bibitem[\protect\citeauthoryear{T. Guver et al.}{2010}]{Guver:1}
%T. Guver {\it et al.}, ApJ {\bf712}, 964 (2010).

%\bibitem[\protect\citeauthoryear{T. Guver et al.}{2010}]{Guver:2}
%T. Guver {\it et al.}, ApJ {\bf719}, 1807 (2010).

%\bibitem[\protect\citeauthoryear{M. A. Perez-Garcia et al.}{2010}]{Stone:1}
%M.A. Perez-Garcia, J. Silk and J. R. Stone, 2010, arXiv:1007.1421v2

%\bibitem[\protect\citeauthoryear{B.K. Harrison et al.}{1965}]{Harrison:1}
%B. K. Harrison {\it et al.}, Gravitation Theory and Gravitational Collapse, 1965 
%(Chicago: Chicago Univ. Press)

%\bibitem[\protect\citeauthoryear{W.D. Langer et al.}{1969}]{Larger:1}
%W. D. Larger and A. G. W. Cameron, Ap\&SS {\bf5}, 213 (1969).

%\bibitem[\protect\citeauthoryear{C. Kouvaris}{2007}]{Kouvaris:1}   
%C. Kouvaris, Phys. Rev. D {\bf 77}, 023006 (2007).

%\bibitem[\protect\citeauthoryear{I. Goldman et al.}{1989}]{Goldman:1}
%I. Goldman and S. Nussinov, Phys. Rev. D {\bf 40}, 10, 3221 (1989).

%\bibitem[\protect\citeauthoryear{J.F. Navarro et al.}{1995}]{Navarro1996b}
%J. F. Navarro, C. S. Frenk and S. D. M. White, 1996b, ApJ, {\bf 462}, 563

%\bibitem[\protect\citeauthoryear{J.F. Navarro et al.}{1995}]{Navarro1997}
%J. F. Navarro, C. S. Frenk and S. D. M. White, 1997, ApJ, {\bf 490}, 493

%\bibitem[\protect\citeauthoryear{B. Moore et al.}{1998}]{Moore1998}
%B. Moore, F. Governato, T. Quinn, J. Statal and G. Lake, 1998, ApJ, {\bf 499}, L5

%\bibitem[\protect\citeauthoryear{B. Moore et al.}{1999}]{Moore1999}
%B. Moore, T. Quinn, F. Governato, J. Statal and G. Lake, 1999, MNRAS, {\bf 310}, 1147

%\bibitem[\protect\citeauthoryear{F. Alcock et al.}{2001}]{Alcock2001}
%F. Alcock {\it et al.}, 2001, ApJ, {\bf 560}, L169

%\bibitem[\protect\citeauthoryear{F. Alcock et al.}{2000}]{Alcock2000}
%F. Alcock {\it et al.}, ApJ {\bf 542}, 281 (2000).

%\bibitem[\protect\citeauthoryear{G. Narain et al.}{2006}]{Narain2006}
%G. Narain, J. Schaffner-Bielich and I. N. Mishustin, PRD {\bf 74}, 063003 (2006).

%\bibitem[\protect\citeauthoryear{G. Ogiya et al.}{2011}]{Ogiya2011}
%G. Ogiya, and M. Mori, 2011, arXiv:1106.2864v1 [astro-ph.CO]

%%%%%%%%%%%%%%%%%%%%%%%%%%%%%%%%%%%%%%%%%%%%%%%%%%%%%%%%%%%%%%%%%%%%%%5
%Complete set
%\bibitem[\protect\citeauthoryear{F. Zwicky}{1933}]{Zwicky:1}
%F. Zwicky, 1933, Astrophys. J., 6, 110 

%\bibitem[\protect\citeauthoryear{J.A. Wheeler}{1961}]{Wheeler:1}
%J.A. Wheeler, 1961, Rev. Mod. Phys., 33, 1, 63

%\bibitem[\protect\citeauthoryear{S. Chandrasekhar}{1964}]{Chandrasekhar:1}
%S. Chandrasekhar, 1964, ApJ, 140, 417

%\bibitem[\protect\citeauthoryear{K.S. Thorne et al.}{1967}]{Thorne:1}
%K.S. Thorne and A. Campolattaro, 1967, ApJ, 149, 591

%\bibitem[\protect\citeauthoryear{W. Laner et al.}{1969}]{Laner:1}
%W. Laner and A. Cameron, 1969, Astrophys. and Space Sci., 5, 213


%\bibitem[\protect\citeauthoryear{R.H. Miller et al.}{1970}]{Miller:1}
%R.H. Miller, K.H. Prendergast and W.J. Quirk, 1970, ApJ, {\bf161}, 903


%\bibitem[\protect\citeauthoryear{F. Hohl}{1971}]{Hohl:1}
%F. Hohl, 1971, Astrophys. J., {\bf168}, 343

%\bibitem[\protect\citeauthoryear{Y.B. Zeldovich et al.}{1971}]{Zeldovich:1}
%Y.B. Zeldovich and I. Novikov, 1971, Star and Relativity, Univ. of Chicago
%Press, Illi.


%\bibitem[\protect\citeauthoryear{J.P. Ostriker et al.}{1973}]{Ostriker:1}
%J.P. Ostriker and P.J.E. Peebles, 1973, Astrophys. J., {\bf186}, 467

%\bibitem[\protect\citeauthoryear{C.W. Misner et al.}{1973}]{MTW:1}
%C.W. Misner, K.S. Throne and K.S. Wheeler, 1973, Gravitation. Freeman \& Co., San Francisco

%\bibitem[\protect\citeauthoryear{J.B. Hartle et al.}{1974}]{Hartle:1}
%J.B. Hartle, R.F. Sawyer and D.J. Scalapino, 1975, ApJ, 199, 471

%\bibitem[\protect\citeauthoryear{G. Chanmugam}{1977}]{Chanmugam:1}
%G. Chanmugam, 1977, ApJ, 217, 799

%\bibitem[\protect\citeauthoryear{M. Milgrom}{1983}]{Milgrom:1}
%M. Milgrom, 1983, Astrophys. J., 270, 365

%\bibitem[\protect\citeauthoryear{S. Detweiler et al.}{1984}]{Detweiler:1}
%S. Detweiler and L. Lindblom, 1984, ApJ, 292, 12

%\bibitem[\protect\citeauthoryear{G. Efstathiou et al.}{1984}]{Efstathiou:1}
%G. Efstathiou and J.R. Bond, 1986, Phil. Trans. Roy. Soc. London, Series A, Math. Phys. Sci, {\bf320}, 585

%\bibitem[\protect\citeauthoryear{Alcock et al.}{1984}]{Alcock:1}
%C. Alcock, E. Farhi and A. Olinto, 1986, ApJ, {\bf310}, 261

%\bibitem[\protect\citeauthoryear{I. Goldman et al.}{1989}]{Goldman:1}
%I. Goldman and S. Nussinov, 1989, Phys. Rev. D, 40, 10, 3221

%\bibitem[\protect\citeauthoryear{S.C. Pandey et al.}{1991}]{Pandey:1}
%S.C. Pandey, M.C. Durgapal and AK. Pande, 1991, Ap \& SS., 180, 75

%\bibitem[\protect\citeauthoryear{K.D. Kokkotas}{1994}]{Kokkotas:0}
%K.D. Kokkotas, 1994, Mon. Not. R. Astron. Soc., 268, 1015

%\bibitem[\protect\citeauthoryear{U. Lee}{1995}]{Lee:1}
%U. Lee, 1995, Astron.Astrophys, {\bf303}, 515

%\bibitem[\protect\citeauthoryear{J.F. Navarro et al.}{1995}]{Navarro:1}
%J.F. Navarro and C.S. Frenk and S.D.M. White, 1995, ApJ, 463, 563

%\bibitem[\protect\citeauthoryear{N. Andersson et al.}{1996}]{Andersson:1}
%N. Andersson, K.D. Kokkotas and B.F. Schutz, 1996, Mon. Not. Roy. Astron. Soc., 280, 1230

%\bibitem[\protect\citeauthoryear{G.L. Comer et al.}{1999}]{Lin:1}
%G.L. Comer, D. Langlois and L.-M. Lin, 1999, Phys. Rev. D, {\bf60}, 104025

%\bibitem[\protect\citeauthoryear{H. Beyer et al.}{1999}]{Beyer:1}
%H. Beyer and K.D. Kokkotas, 1999, Mon. Not. Roy. Astron. Soc., 308, 745

%\bibitem[\protect\citeauthoryear{K.D. Kokkotas et al.}{1999}]{Kokkotas:1}
%K.D. Kokkotas and B.G. Schmidt, 1999, LivingRev. Rel., 2, 2

%\bibitem[\protect\citeauthoryear{G.L. Comer et al.}{2001}]{Lin:2}
%N. Andersson, G.L. Comer, 2001, Class. Quantum Grav., {\bf18}, 969-1002

%\bibitem[\protect\citeauthoryear{K.D. Kokkotas et al.}{2001}]{Kokkotas:2}
%K.D. Kokkotas and N. Andersson, 2001, arXiv:gr-qc/0109054v1

%\bibitem[\protect\citeauthoryear{K.D. Kokkotas et al.}{2001}]{Kokkotas:3}
%K.D. Kokkotas and J. Ruoff, 2001, Astron.Astrophys. 366, 565 

%\bibitem[\protect\citeauthoryear{R. Prix et al.}{2001}]{Prix:1}
%R. Prix, G.L. Comer and N. Andersson, 2001, Astron. Astrophys., {\bf381}, 178

%\bibitem[\protect\citeauthoryear{N. Andersson et al.}{2002}]{Andersson:2}   
%N. Andersson, G.L. Comer, D. Langlois, 2002, Phys. Rev. D, 66, 104002

%\bibitem[\protect\citeauthoryear{M. Bejger et al.}{2002}]{Haensel:1}
%M. Bejger, P. Haensel, 2002, A \& A, {\bf396}, 917-921

%\bibitem[\protect\citeauthoryear{K.D. Kokkotas et al.}{2002}]{Kokkotas:4}
%K.D. Kokkotas and J. Ruoff, 2002, arXiv:gr-qc/0212105v1

%\bibitem[\protect\citeauthoryear{R. Prix et al.}{2002}]{Prix:2}
%R. Prix and M. Rieutord, 2002, Astron. Astrophys., 393, 949

%\bibitem[\protect\citeauthoryear{P.K. Sahu et al.}{2002}]{Sahu:1} 
%P.K. Sahu, G.F. Burgio and M. Baldo, 2002, Astrophys. J., 566, L89          

%\bibitem[\protect\citeauthoryear{A. Stavridis et al.}{2004}]{Stavridis:1}
%A. Stavridis and K.D. Kokkotas, 2004, Int. J. Mod. Phys., D14, 543

%\bibitem[\protect\citeauthoryear{J.L. Feng}{2005}]{Feng:1}
%J.L. Feng, 2005, J. Phys, G32, R1

%\bibitem[\protect\citeauthoryear{R.J. Smith et al.}{2005}]{Smith:1}
%R.J. Smith et al., 2005, ApJ, {\bf625}, L103

%\bibitem[\protect\citeauthoryear{M. Alford et al.}{2006}]{Alford:1}
%M. Alford, D. Blaschke, A. Drago, T. Klahn, G. Pagliara and J. Schaffner-Bielich, 2002, nature, 445, E7

%\bibitem[\protect\citeauthoryear{R. Aquilano}{2006}]{Aquilano:1}
%R. Aquilano, 2006, Astrop. Phys., 26. 64

%\bibitem[\protect\citeauthoryear{G. Bertone et al.}{2007}]{Bertone:1}
%G. Bertone and M. Fairbairn, 2007, Phys. Rev. D, {\bf77}. 043515

%\bibitem[\protect\citeauthoryear{C. Kouvaris}{2007}]{Kouvaris:1}
%C. Kouvaris, 2007, Phys. Rev. D, 77, 023006 

%\bibitem[\protect\citeauthoryear{R. Massey et al.}{2007}]{Massey:1}
%R. Massey et al., 2007, nature, {\bf445}, 286

%\bibitem[\protect\citeauthoryear{D.N. Spergel et al.}{2007}]{Spergel:1}
%D.N. Spergel et al., 2007, ApJ Suppl., 170, 377

%\bibitem[\protect\citeauthoryear{C. Ishizuka et al.}{2008}]{Ishizuka:1}
%C. Ishizuka, A. Ohnishi, K. Tsubakihara, K. Sumiyoshi and S. Yamada, 2008, JPhys, G35, 085201

%\bibitem[\protect\citeauthoryear{Kholopov et al.}{2008}]{Kouvaris:2}
%M.Y. Kholopov and C. Kouvaris, 2008, Phys. Rev. D, {\bf77}, 065002

%\bibitem[\protect\citeauthoryear{Petriello et al.}{2008}]{Petriello:1}
%F. Petriello and K. M. Zurek, 2008, JHEP, {\bf0809}, 047


%\bibitem[\protect\citeauthoryear{P. Ciarcelluti}{2009}]{Ciarcelluti:1}
%P. Ciarcelluti, 2009, arXiv:0911.3592v1

%\bibitem[\protect\citeauthoryear{J. Diemand et al.}{2009}]{Diemand:1}
%J. Diemand and B. Moore, 2009, arXiv:0906.4340v1

%\bibitem[\protect\citeauthoryear{K. Freese et al.}{2009}]{Freese:1}
%K. Freese, D. Spolyar, P. Bodenheimer, P.  Gondolo, 2009, arXiv:0903.0101v1

%\bibitem[\protect\citeauthoryear{M. Gabler et al.}{2009}]{Gabler:1}
%M. Gabler, U. Sperhake and N. Andersson, 2009, Phys. Rev. D, 80, 064012

%\bibitem[\protect\citeauthoryear{D.E. Kaplan et al.}{2009}]{Kaplan:1} 
%D.E. Kaplan, M.A. Luty, K.M. Zurek, 2009, Phys. Rev. D, {\bf79}, 115016

%\bibitem[\protect\citeauthoryear{F. Saudin et al.}{2009}]{Ciarcelluti:2}
%F. Saudin, P. Ciarcelluti, 2009, Astropart. Phys., {\bf32}, 5, 278  

%\bibitem[\protect\citeauthoryear{F. Ozel et al.}{2009}]{Ozel:1}
%F. Ozel, T. Guver and D. Psaltis, 2009, ApJ, {\bf693}, 1775

%\bibitem[\protect\citeauthoryear{P. Gondolo et al.}{2009}]{Gondolo:1}
%C. Savage, K. Freese, P. Gondolo and D. Spolyar, 2009, JCAP, {\bf0909}, 036

%\bibitem[\protect\citeauthoryear{P. Gondolo et al.}{2009}]{Gondolo:2}
%C. Savage, K. Freese, P. Gondolo and D. Spolyar, 2009, JCAP, {\bf0904}, 010

%\bibitem[\protect\citeauthoryear{H. Umeda et al.}{2009}]{Umeda:1}
%H. Umeda, N. Yoshida, K. Nomoto, S. Tsuruta, M. Sasaki, T. Ohkubo, 2009, JCAP, 08, 024

%\bibitem[\protect\citeauthoryear{C. E. Aalseth et al.}{2010}]{Aalseth:1}
%C. E. Aalseth et al (the CoGeNT Collaboration), 2010, arXiv:1002.4703

%\bibitem[\protect\citeauthoryear{S. Ando et al.}{2010}]{Kusenko:1}
%S. Ando and A. Kusenko, 2010, Phys. Rev. Lett., 81, 113006 

%\bibitem[\protect\citeauthoryear{V. Berezinsky et al.}{2010}]{Berezinsky:1}
%V. Berezinsky, V. Dokuchaev, Y. Eroshenko, M. Kachelriess and  M.A. Solberg, 2010, Phys. Rev. D, 81, 103529

%\bibitem[\protect\citeauthoryear{R. Bernabei et al.}{2010}]{Bernabei:1}
%R. Bernabei, P. Belli, F. Cappella et al., 2010, Eur. Phys. J., {\bf C67}, 39-49

%\bibitem[\protect\citeauthoryear{P. Ciarcelluti et al.}{2010}]{Ciarcelluti:3}
%P. Ciarcelluti and F. Sandin, 2010, arXiv:1005.0857v1

%\bibitem[\protect\citeauthoryear{D.T. Cumberbatch et al.}{2010}]{Cumberbatch:1}
%D.T. Cumberbatch, J.A. Guzik, J. Silk, L.S. Watson and S.M. West, 2010, Phys. Rev. D., 82, 103503

%\bibitem[\protect\citeauthoryear{P.B. Demorest et al.}{2010}]{Demorest:1}
%P. B. Demorest, T. Pennucci, S. M. Ransom, M. S. E. Roberts and J. W. T. Hessels, 2010, nature, {\bf467}, 1081


%\bibitem[\protect\citeauthoryear{J.L. Feng et al.}{2010}]{Feng:2}
%J.L. Feng, 2010, arXiv:1003.0904v2

%\bibitem[\protect\citeauthoryear{M.T. Frandsen et al.}{2010}]{Frandsen:1}
%M.T. Frandsen and S. Sarkar, 2010, Phys. Rev. Lett., 105, 011301

%\bibitem[\protect\citeauthoryear{D. Gonzalez et al.}{2010}]{Gonzalez:1}
%D. Gonzalez and A. Reisenegger, 2010, arXiv:1005.5699v3

%\bibitem[\protect\citeauthoryear{T. Guver et al.}{2010}]{Guver:1}  
%T. Guver et al., 2010, ApJ, {\bf712}, 964  

%\bibitem[\protect\citeauthoryear{T. Guver et al.}{2010}]{Guver:2}
%T. Guver et al., 2010, ApJ, {\bf719}, 1807

%\bibitem[\protect\citeauthoryear{X. Hernandez et al.}{2010}]{Hernandez:1}
%X. Hernandez and W.H. Lee , 2010, arXiv:1002.0553v1

%\bibitem[\protect\citeauthoryear{D. Hooper et al.}{2010}]{Hooper:1}
%D. Hooper et al., 2010, prd, {\bf82}, 123509

%\bibitem[\protect\citeauthoryear{L. Iorio}{2010}]{Iorio:1}
%L. Iorio, 2010, JCAP, 1005, 018

%\bibitem[\protect\citeauthoryear{L. Iorio}{2010}]{Iorio:2}
%L. Iorio, 2010, arXiv:1005.5078v2

%\bibitem[\protect\citeauthoryear{C. Kouvaris et al.}{2010}]{Kouvaris:3}
%C. Kouvaris and P. Tinyakv, 2010, arXiv:1004.0586v1

%\bibitem[\protect\citeauthoryear{A. Lavallaz et al.}{2010}]{Lavallaz:1}
%A. de Lavallaz and M. Fairbairn, 2010, arXiv:1004.0629v1

%\bibitem[\protect\citeauthoryear{M. A. Perez-Garcia et al.}{2010}]{Stone:1}
%M.A. Perez-Garcia, J. Silk and J. R. Stone, 2010, arXiv:1007.1421v2

%\bibitem[\protect\citeauthoryear{E. Ripamonti et al.}{2010}]{Ripamonti:1}
%E. Ripamonti, F. Iocco, A. Ferrara, R. Schneider, A. Bressan and P. Marigo, 2010, arXiv:1003.0676v1

%\bibitem[\protect\citeauthoryear{M. Roos}{2010}]{Roos:1}
%M. Roos, 2010, arXiv:1001.0316v2
 
%\bibitem[\protect\citeauthoryear{M. Taoso et al.}{2010}]{Taoso:1}
%M. Taoso, F. Iocco, G. Meynet, G. Bertone and P. Eggenberger, 2010, arXiv:1005.5711v2




\end{thebibliography}

\end{document}